\documentclass[12pt]{article}
\usepackage{amsmath,amsfonts,amssymb}
\usepackage[english]{babel}
\usepackage{graphicx}
\usepackage{color}
\usepackage{xtab}
\newcommand{\be}{\begin{equation}}
\newcommand{\ee}{\end{equation}}
\newcommand{\bea}{\begin{eqnarray}}
\newcommand{\eea}{\end{eqnarray}}
\newcommand{\vr}{\varepsilon}

\begin{document}
\pagestyle{empty}

\vfill 
 
 \vspace{30mm}

\begin{center}

~~\\
~~\\
~~\\

{\Large  \textbf{\textsf{ About the Non Relativistic Structure of the  AdS/CFT Superalgebras}}}

\vspace{10mm}

{\large  A. Sciarrino$^{ab}$, P. Sorba$^{c}$}

\vspace{10mm}

\emph{$^a$ Dipartimento di Scienze Fisiche, Universit{\`a} di Napoli
``Federico II''}

\emph{$^b$ I.N.F.N., Sezione di Napoli}

\emph{Complesso Universitario di Monte S. Angelo}

\emph{Via Cinthia, I-80126 Napoli, Italy}

\vspace*{2mm}

\emph{$^c$ $LAPTH^1$ , Universit\'e de Savoie, CNRS \\
9 Chemin de Bellevue, BP110, F-74941 Annecy-le-Vieux}

\vspace{12mm}

\end{center}

\vspace{7mm}

\begin{abstract}
  The property of the conformal algebra to contain the Schr\"odinger algebra in one less space dimension is extended to the supersymmetric case. More precisely, we determine the counterpart of any field theory admissible super conformal algebra. Even if each type of superalgebra provides a different solution, its basis decomposition into two copies of the super  Schr\"odinger  algebra, differing only by their super Heisenberg part, remains valid in all the cases, so generalizing a feature already observed in the non supersymmetric conformal case.     
\end{abstract}

\vfill
PACS number:  02.20.-a, 11.30.Pb, 12.60.Jv, 04.65.+e.

\vspace*{2mm}
Keywords:  Schr\"odinger (super)algebra; (super)conformal algebra; Heisenberg (super)algebra;

real form of superalgebras; AdS/CFT correspondence.
\vfill

\rightline{DSF-09/2010}
\rightline{LAPTH-028/10}

\vspace*{3mm}
\hrule
\vspace*{2mm}
\noindent
\emph{E-mail:} \texttt{sciarrino@na.infn.it},  \texttt{sorba@lapp.in2p3.fr}

\noindent
$^1$ Laboratoire de Physique Th\'eorique d'Annecy-le-Vieux, UMR 5108

\newpage
\phantom{blankpage}
\newpage
\pagestyle{plain}
\setcounter{page}{1}
\baselineskip=16pt

\section{Introduction}

Soon after the discovery of the Schr\"odinger  algebra, defined as the maximal kinematical symmetry of the  Schr\"odinger equation  \cite{Niederer}, (see also  \cite{Hagen}), it was realized that such an 
algebra, defined in $d$ space and 1 time dimensions, is  embedded in the conformal algebra acting on the Minkowsky space with ($d+1$) space and 1 time dimensions \cite{BPS}. Actually, using a well adapted basis for the conformal algebra, two conjugated   Schr\"odinger algebras ${\it \widetilde{Sch}(d-1,1)},$ with a common part constituted by the rotations and Òconformal transformationsÓ, can be recognized in {\it SO(d+1,2)} in a way analogous to the decomposition of the conformal algebra into two conjugate Poincar\'e ones. We remind that in this last case the conformal algebra can be seen as the direct sum of the Lorentz algebra plus a dilatation, to which can be added, on one side,  the ($d+1$) translations and, on the other side, the ($d+1$) special conformal transformations, each set constituting a ÒPoincar\'e-likeÓ algebra.

In these last  years, there has been a lot of activity in studying, in the framework of AdS/CFT correspondence, the string theory embedding of a spacetime with non relativistic  Schr\"odinger symmetry \cite{MMT}. Indeed, it is first tempting, due to the difficulty of solving the general problem, to look for a limit - actually a non relativistic one -  of the AdS superstring. This has led to the study of some gravity backgrounds, candidates to be gravity dual of conformal quantum mechanical systems. But, even more interestingly, such an approach is particularly useful in the context of several condensed matter systems \cite{Son},  \cite{YSon}. Indeed, the microscopic description of many condensed matter systems in the vicinity of a quantum critical point is certainly not relativistic, but exhibits conformal invariance \cite{Hartnoll}. Let us emphasize, at this point, that the treatment of strongly anisotropic critical systems with the help of   Schr\"odinger invariance had already been proposed, about fifteen years ago, in  \cite{Henkel}.

Such an interest in the  Schr\"odinger invariance naturally raises the question of its supersymmetric extension, and more precisely of the embedding of such a supersymmetric analog in the superconformal algebras.
But then, it is natural to ask whether an intrinsic definition of the super  Schr\"odinger algebra can be given, the  Schr\"odinger algebra having been introduced as the maximal kinematical invariance algebra of the  Schr\"odinger equation. A first answer to this question has been 
given in \cite{GGT} where the non relativistic spin $1/2$ particle action is constructed. At about
the same period, an extended superconformal Galilean algebra, which slightly differs from the one of \cite{GGT}, was proposed in the context of the non relativistic limit of N=2 supersymmetric Chern-Simons matter systems - and therefore in the special case of $d=2$ space dimensions, see \cite{LLM}. Such a difference was clarified by the authors of  \cite{DH} who performed a more direct algebraic construction of supersymmetric extensions of the  Schr\"odinger algebra: we will come back to this approach in Section 3. 

The list of admissible superconformal algebras is known since long, and can be easily selected from the simple super algebras in Nahm classification of manifest supersymmetries \cite{Nahm}.  They are: 
 the simple unitary superalgebras  ${\it SU(2,2/N)}$, the orthosymplectic ones  ${\it Osp(8^*/N)}$, with $N$ even, 
 \newpage
 \noindent and ${\it Osp(M/4,R)}$ and the real form commonly denoted \footnote{The four real forms of {\it F(4)} are often denoted as  ${\it F(4;p)}$  with p = 0,1,2,3,  the complete bosonic part of ${\it F(4;p)}$ being {\it Sl(2,R)} $ \oplus $ {\it SO(p,7-p)} for p = 0, 3 and {\it SU(2)} $ \oplus $ {\it SO(p,7-p)} for p = 1, 2. We note at this point a misprint in Table.3.75 of \cite{FSS}.}
${\it F(4;2)}$ with bosonic part {\it SU(2) $ \oplus $ SO(5,2)} of the exceptional super-algebra {\it F(4)}. One obviously recovers in {\it pSU(2,2/4}) the supersymmetry algebra of the $10$-dimensional Type II supergravity, and in {\it Osp(8/4)} the one of the $11$-dimensional M-theory with corresponding space-time manifolds $AdS_5 \times S^5$ and $AdS_7 \times S^4$ respectively, while {\it Osp(8/4,R)} is related to the M-theory with space-time manifold $AdS_4  \times S^7$. Finally the exceptional solution has also been considered
and the $AdS(6)/CFT(5)$  correspondence investigated for a {\it F(4;2)} supergravity theory \cite{DFV}.

The purpose of this paper is to determine, for each admissible superconformal symmetry, its associated super Schr\"odinger algebra, natural extension of the Schr\"odinger part contained in the conformal algebra. A special care will be brought to present each algebra in an explicit form, that is as a semi-direct sum of a ÒreductiveÓ part, i.e. a direct sum of simple (super)algebras, eventually with {\it U(1)} factors, acting on a super Heisenberg part. Moreover, it will be shown that each such a super conformal algebra contains a couple of super Schr\"odinger algebras, natural extension of the couple of Schr\"odinger algebras embedded in its bosonic conformal part.
At this point, we must mention \cite{SY}, where the problem of finding a super Schr\"odinger algebra in each of the three super algebras {\it pSU(2,2/4)}, {\it Osp(8*/4)} and {\it Osp(6,2/4)} has recently been considered using adequate projections on the spinorial fermionic sectors. We agree with the presented results, but would like to insist on the more general aspects of our approach, which, as just mentioned above, proposes an as complete as possible characterization of super Schr\"odinger algebras and also includes the {\it F(4,2)} case.

The plan of our paper is as follows. We start, in Section 2, by a reminder on the Schr\"odinger algebra showing up as a  subalgebra of the conformal one in one more space dimensions and some comments on the special position of this subalgebra inside the conformal symmetry. We comment, in Section 3, on the super Schr\"odinger symmetries already discovered in physical situations and the attempt to recognize them in a supersymplectic framework, as proposed in  \cite{DH}; our method of determination of the super Schr\"odinger algebras inside super conformal ones is also rapidly summarized. Then, a separate section is devoted to the construction of the Schr\"odinger counterpart for each family of superconformal algebras:  that is the unitary {\it SU(2,2/N)} for $N$ different from $4$, and {\it pSU(2,2/4)}  
 superalgebras with conformal symmetry in $3+1$ dimensions (Section 4); the orthosymplectic
{\it Osp(N/4,R)}  in $2+1$ dim. (Section 5) and  the orthosymplectic {\it Osp(6,2/2N)} ones in 5+1 
dim. (Section 6);  finally the exceptional {\it  F(4;2)} algebra in $5+1$ space-time dim. (Section 7). A comparison between the different types of obtained Schr\"odinger superalgebras (Section 8) concludes the paper.

\section{Schr\"odinger algebra inside Conformal Algebra}

As firstly remarked in  \cite{BPS}, an adequate choice of basis for the conformal algebra acting on the Minkowski space {\em M(d,1)} allows to identify - up to an isomorphism - the Schr\"odinger algebra
${\it \widetilde{Sch}(d-1)}$, i.e. in ($d-1$) space dimensions and in $1$ time dimension, as a subalgebra of 
{\it SO(d+1,2)}. Actually, two copies of ${\it \widetilde{Sch}(d-1)}$, one conjugate to the other and sharing a semi-simple common part, can be used for recontructing, up to one dimensional generator, the conformal algebra. We hereafter explicite the (most common) case $d=3$, the generalisation to any $d$ being straightforward. So, let us consider the {\it SO(4,2)} algebra generated by the 15 skew-symmetric elements $M_{\mu \nu}$  ($\mu, \nu= 0,0',1,2,3,4$) such that:
\be
[M_{\mu \nu}, \, M_{\rho \sigma}] = g_{\mu \sigma} M_{\nu \rho} +  g_{\nu\rho} M_{\mu \sigma} - g_{\nu   \sigma} M_{\mu \rho}  - g_{\mu \rho} M_{\nu \sigma}  \label{eq:skew}
\ee
The metric $g_{\mu \nu}$ satisfying
\be
g_{\mu \nu} = 0 \;\;\;\;  \mu \neq \nu, \;\;\;\;\;\;\;\;  g_{\mu \mu} = 1  \;\;\;\; \; \mu = 0,0', 
\;\;\;\;\;\;\;\;  g_{\mu \mu} = -1  \;\;\;\;  \mu = 1,2,3,4
\ee
We note that the $M_{\mu \nu}$, where  $\mu, \nu= 0,1,2,3,$ can be seen as the generators the Lorentz algebra. Then the combinations:
\be
p_{\mu} = M_{0' \mu} + M_{4 \mu} \;\;\;\;\;\;\;\;  \mbox{and } \;\;\;\;\;\;\;\;  c_{\mu} = M_{0' \mu} -  M_{4 \mu}
\;\;\;\;\;\;\;\;  \mu = 0,1,2,3
\ee
correspond to the translations and special conformal transformations, each set forming, with {\it SO(3,1)}, a Poincar\'e-like algebra, and these two-dimensional subalgebras being conjugate in {\it SO(4,2)}.
As a last generator, we have $M_{0' 4}$, acting as a dilatation on the  $p_{\mu}$'s and $c_{\mu}$'s  and commuting with the Lorentz part.  One can first select, in (one of) the Poincar\'e part, the $2$-dimensional extended Galilean algebra  by performing a change of basis,  already used around the $70$'s in the infinite momentum formalism.
Let us define
\bea\;\;\;\;
H & = & \frac{1}{2} \left(p_0 - p_3  \right)  \;\;\;\;\;\;\;\;\;\;\;\;  M = p_0 + p_3   \;\;\;\;\;\;\;\;  J_{3} = M_{12}  \label{eq:HM} \\
K_{a} &  = &  -(M_{0a} + M_{3a})  \;\;\;\;\;\;\;\; \;\;\;\;  P_a = p_a  \;\;\;\;\;\;\;\;\;\;\; a = 1,2   \label{eq:KP}
\eea
satisfying the non vanishing commutation relations:
\bea
&& [J_3, \, P_a ]   =    \vr_{3ab}  P_b  \;\;\;\; \;\;\;\;\;\;\;\;\; \,  [H, \, K_a ] = -   K_a \;\;\;\;\;\;\;\;\;\;\ a,b = 1,2   \\
&& [J_3, \, K_a ]   =   \vr_{3ab}  K_b   \;\;\;\;\;\;\;\;\;\;\;\;  [K_a, \, P_b ] = -   \delta_{ab} \, M  
\eea
To this Gailean algebra  the two generators
\be
C =  \frac{1}{2} \left(c_0 + c_3  \right)  \;\;\;\;\;\;\;\;  \mbox{and } \;\;\;\;\;\;\;\;  D = M_{0' 4} -  M_{03}
\ee
can be added which, together with $H$, defined in eq.(\ref{eq:HM}), close into an algebra isomorphic to

\noindent {\it SU(1,1) $ \cong $ SO(2,1)}
\be
    [C, \, H] = D  \;\;\;\;\; \;\;\;\;\;     [D, \, C] =    2 C \;\;\;\;\; \;\;\;\;\;  [D, \, H] = - 2 H  \label{eq:su11}
    \ee
We note that this  {\it SU(1,1)} algebra commutes with $J_3$ and $M,$  and acts on the {\it H(2)} Heisenberg part generated  by $P_a, K_a$ and $M$ as it follows
\be
    [D, \, P_a] = - P_a \;\;\;\;\; \;\;\;\;\;       [C, \, P_a] =   K_a \;\;\;\;\; \;\;\;\;\;     [D, \, K_a] = K_a 
    \ee
     The 5 dimensional {\it H(2)} Heisenberg together  with the {\it SO(2)} algebra generated by $J_3,$ and the  {\it SU(1,1)} defined in eq.(\ref{eq:su11}) span an $9$ dimensional  {\it SO(4,2)} subalgebra which is isomorphic to the (extended) Schr\"odinger algebra {\it $\widetilde{Sch}(2)$} in 2 space and 1 time dimensions. It can be conveniently seen as the semi-direct sum of the {\it SO(2)} rotation algebra and the
     {\it SU(1,1)} conformal part acting on the  {\it H(2)}  Heisenberg algebra  \footnote{We denote by $\, \rhd \,$ the semi-direct sum.} :
     \be
{\it \widetilde{Sch}(2)} =  [{\it SO(2)} \oplus  {\it SU(1,1)}] \,  \rhd \,  {\it H(2)}  \label{eq:sch1}
\ee
Such an inclusion among algebras can be formally illustrated by rewriting the d'Alembert equation
\be
p^2 \, |0> = 0  \;\;\;\;\; \;\;\;\;\;  p^2  = p_0^2  - p_1^2 - p_2^2  -  p_3^2
\ee
$|0> $ being the state of a relativistic massless particle in the momentum space
\be
\left(H - \frac{p_{\perp}^2}{M} \right) |0>  = 0 \label{eq:KG}
\ee
with $p_{\perp} = \{ p_1, p_2\}$ for states such that $M^2 \, |0> \neq 0$.
We recognize in eq.(\ref{eq:KG}) the Schr\"odinger equation for a free massive non relativistic particle in one less space dimension and invariant under {\it $\widetilde{Sch}(2)$}. Now in the same way a   {\it $\widetilde{Gal}(2)$}  Galilean algebra can be extracted from the Poincar\'e algebra generated by $M_{\mu\nu}$ and $p_{\mu}$, a {\it $\widetilde{Gal}^{*}(2)$} algebra can obviously be obtained from the set $\{M_{\mu\nu},  c_{\mu}\}$. The {\it $ \widetilde{Gal}^{*}(2)$} algebra is conjugate to {\it $\widetilde{Gal}(2)$} in {\it SO(4,2)}  and is simply constructed by adding to the previously introduced rotation  $J_3$ and time-translation $H$ generators the elements
\be
p^*_a  = M_{0a} -  M_{3a}   \;\;\;\;\; \;\;\;\;\;  K^*_a = c_a   \;\;\;\;\; \;\;\;\;\;  M^* = c_0 -  c_3  \;\;\;\;\; \;\;\;\;\; a = 1,2 \label{eq:h*}
\ee
The 5 generators of eq.(\ref{eq:h*}) form a Heisenberg algebra that we denote {\it $ H^{*}(2)$}. Then, keeping the same {\it SU(1,1)} conformal algebra generated by $H,C,D$, one gets a second $2+1$ dimensional 
Schr\"odinger algebra  {\it $ \widetilde{Sch}^*(2)$}:
  \be
{\it \widetilde{Sch}^*(2)} =  [{\it SO(2)} \oplus  {\it SU(1,1)}] \,  \rhd \,  {\it H^*(2)} \label{eq:sch}
\ee
with {\it $\widetilde{Sch}(2)$}  and {\it $\widetilde{Sch}^*(2)$}, conjugate one to the other, differing therefore by their Heisenberg part. Finally, as a vector space (v.s.), one can decompose the algebra {\it SO(4,2)} as follows
\be
{\it SO(4,2)} =_{v.s.}  {\it SO(2)} \oplus  {\it SO(2,1)} \oplus  {\it H(2)} 
\oplus  {\it H^*(2)} \oplus  {\it SO(1,1)}   \label{eq:dec}
\ee
with as a {\it SO(1,1)}  generator
\be
\Lambda = -(M_{0'4} -  M_{03})
\ee
itself acting as a scale transformation on the Heisenberg algebra elements
\bea
&& [\Lambda, \, P_a]   =    P_a  \;\;\;\;\; \;\;\;\;\;\;\;  [\Lambda, \, P^*_a] =-  P^*_a  \nonumber \\
&&[\Lambda, \, K_a]   =    K_a  \;\;\;\;\; \;\;\;\;\;\,  [\Lambda, \, K^*_a] =-  K^*_a  \nonumber \\
&& [\Lambda, \, M]  =   2  M  \;\;\;\;\; \;\;\;\;\;  [\Lambda, \, M^*] =- 2  M^*  \label{eq:su11-2}
 \eea 
 in a way completely analogous to the dilatation  $M_{0'4} $ acting on the translations $P_{\mu}$ and special conformal transformations $C_{\mu}$.
 Let us note that the three generators  $\Lambda, M,  M^*$ close into a {\it SU(1,1)} $\cong $ {\it SO(2,1)} algebra, see eq.(\ref{eq:su11-2}) and
 \be
 [M, \, M^*] =- 4 \Lambda
 \ee
 this {\it SU(1,1)} algebra commuting with the conformal {\it SU(1,1)} above defined in eq.(\ref{eq:su11}) , as well as with the rotation 
 {\it SO(2)}. So, we recognize the embedding
 \be
{\it SO(4,2)} \supset  {\it SO(2)} \oplus  {\it SO(2,2)} =   {\it SO(2)} \oplus  {\it SO(2,1) }\oplus  {\it SO(2,1)}  \label{eq:cal2}
\ee
The generalisation to $d \neq 3$ of the decomposition eq.(\ref{eq:cal2}) is straightforward and, in particular, eq.(\ref{eq:sch}) and eq.(\ref{eq:cal2}) become, respectively,
\bea
{\it \widetilde{Sch}(d-1)} & = &  ({\it SO(d-1)} \oplus  {\it SO(2,1)}) \,  \rhd \,  {\it H(d-1)} \\
{\it SO(d+1,2)} & \supset  & {\it SO(d-1)} \oplus  {\it SO(2,2)} =   {\it SO(d-1)} \oplus  {\it SO(2,1) }\oplus  {\it SO(2,1)}  
\eea
Considering the algebras as vector spaces we can write, generalising eq.(\ref{eq:dec}):
\be
{\it SO(d+1,2)} =_{v.s.} {\it SO(d-1)} \oplus  {\it SO(2,1)} \oplus  {\it H(d-1)} \oplus  {\it H^*(d-1)}  \oplus  {\it SO(1,1)} 
\ee
As a last remark, let us note that the Schr\"odinger algebra ${\it \widetilde{Sch}(d-1)}$  can be seen as the stabilizer of the $M$ generator -i.e. the set of elements commuting with $M$- in the conformal {\it SO(d+1,2)}  algebra.

\section{Schr\"odinger algebra and Supersymmetry}

Supersymmetric extensions of the Schr\"odinger algebra have been considered in different ways, leading to two types of (super)-symmetry algebras.  Superalgebras of the first family - that we will call of the  ``orthosymplectic type'' - hold for any (integer) $N$ supersymmetries and any $d$ space dimension, while those of the second one - that we will call of the  ``unitary type'' - work for any $N$, but only in the case of $d=2$ space dimensions. In both cases, the time dimension is 1. Firstly exhibited in the study of the supersymmetric harmonic oscillator \cite{BH}, in \cite{GGT} the orthosymplectic type appeared as the symmetry of the non-relativistic spin $1/2$ action. Soon later the unitary type was formed by the authors  of \cite{LLM} as the symmetry of the ($d=2$) non-relativistic Chern-Simons matter systems
extending in this way the gr symmetry before discovered in \cite{J} for this theory.
Actually an elegant way to recognize, in the same framework, these two types of symmetry algebras has been proposed in \cite{DH} where a geometrical symplectic approach is used. Indeed, the structure of the - extended - Schr\"odinger algebra ${\it \widetilde{Sch}(d)}$ may suggest to consider ${\it \widetilde{Sch}(d)}$  as a subalgebra of the - central extended - algebra of the inhomogeneous symplectic transformations 
${\it \widetilde{ISp}(2d)},$ i.e. semi-direct sum of the symplectic algebra {\it Sp(2d,R)} \footnote{In our notation {\it Sp(2d,R)} is the symplectic algebra of rank $d$.} acting on the Heisenberg part {\it H(d)} 
\be
{\it \widetilde{ISp}(2d)} \equiv  {\it Sp(2d,R)} \, \rhd \, {\it H(d)} 
\ee
As shown in  \cite{DH} the natural supersymmetric extension is ${\it \widetilde{IOsp}(N/2d)}$ defined as
\be
{\it \widetilde{IOsp}(N/2d)} \equiv {\it Osp(N/2d,R)} \, \rhd \, {\it SH(d/N)} 
\ee
where {\it Osp(N/2d,R)}  is a natural supersymmetric extension of {\it Sp(2d,R)}  and {\it SH(d/N)}  is defined as the super-Heisenberg algebra obtained from {\it H(d)}  by adding the (fermionic) super-translations generators $\Xi^j_a$ ($a = 1,\ldots,d; \, j = 1,\ldots,N$) which commute with all the 
{\it H(d)} generators
\be
[P_a, \, \Xi^j_b] = [K_a, \, \Xi^j_b]  = [M, \, \Xi^j_b]  = 0
\ee
and satisfy
\be
[\Xi^i_a,  \Xi^j_b] = \delta^{ij} \, \delta_{ab} \, M \;\;\;\; \;\;\;\; a,b = 1,\ldots,d; \, i,j = 1,\ldots,N
\ee
Then the next and final step stands in looking for supersymmetric extensions of  {\it$ \widetilde{Sch}(d)$} in 
{\it $\widetilde{IOsp}(N/2d)$}, via the canonical contact structure on $R^{2d+1}$ suitably extended with the generators $\xi^{j}$  ($j = 1,\ldots,N$) of the Grassmann algebra  ${\bf \wedge R^N}$. As a result, two families of Schr\"odinger superalgebra has been proposed, each of them valid for any positive (or null) integer value of $N$:
\begin{itemize}
\item algebras of the  ``orthosymplectic type'' 
\be
\left[ {\it SO(d)} \, \oplus \, {\it Osp(N/2,R)} \right] \, \rhd \,  {\it SH(d/N)} 
\label{eq:32}
\ee
valid for any value of the positive integer $d$.  {\it SO(d)} is the rotation algebra acting on the translations, super-translations and Galilean boosts. The {\it Sp(2,R)} algebra commuting with {\it SO(N)} in the bosonic sector of  {\it Osp(N/2,R)} is actually the ``conformal" part, generated by $H,C,D$, of the Schr\"odinger algebra  defined in Section 2. It is this kind of Schr\"odinger superalgebra  which shows up in \cite{GGT} for $N=1$ and in \cite{BH}  for $N=2$.
\item  for the special value $d = 2$, another kind of superalgebra has been detected. From the set of
commutation and anti-commutation relations given in \cite{DH}, one can  recognize the
algebra:
\be
             \left[ {\it SO(2)} \, \oplus \,            {\it SU(N/1,1)}\right] \, \rhd  \, {\it SH(2/N)}
                             \label{eq:33}
                             \ee
where the $R$-symmetry standing in the {\it SU(N/1,1)}  bosonic sector is now a {\it SU(N)}  algebra. 
It might be useful to note the doubling in the  number of fermionic generators in going from  {\it Osp(N/2,R)} to  {\it SU(N/1,1)}. 
 At this point let us mention that  ${\it SO(2)}  \oplus {\it SU(N/1,1)}$  is determined as a subalgebra \footnote{See also \cite{DEGKS} for a study of the maximal embeddings: {\it Osp(2m/2n,R)}  $ \supset $ {\it SU(m/p,q)} $\oplus$ {\it U(1)} with $p+q=n; \,  p,q \geq 0$.} of {\it Osp(2N/2,R)}. 
  It is a superalgebra of this kind which has been discovered in \cite{LLM}. We will, rather naturally, denote such superalgebras as Schr\"odinger superalgebras of the unitary type.
\end{itemize}
Let us emphasize that such orthosymplectic - eq.(\ref{eq:32}) - as well as unitary  -  eq.(\ref{eq:33}) -  structures  will show up  as super  counterparts of the  superconformal {\it  Osp(N/4,R)}  and {\it SU(2, 2/N)}  algebras\footnote{As it will be seen in Section 4, the case $N=4$ presents a peculiarity: then starting from {\it pSU(2,2/4)} the  Schr\"odinger symmetry is simply ${\it SU(4/1,1)} \rhd  {\it SH(2/4)}$, i.e. the extra {\it SO(2)} of eq.(\ref{eq:33}) is no more present.} respectively: see Sections 4 and 5 below. But slightly different configurations will be detected in {\it Osp(6,2/2N)} and in {\it  F(4;2)}  superalgebras, as explicited in Sections 6 and 7. 

Obviously, the method for determining the  symmetries used in \cite{DH} is not adapted to our problem. As remarked in the previous section, at the bosonic level the Schr\"odinger algebra is simply the stabilizer in the conformal algebra of the mass generator $M$. It is this property that we plan to extend to the supersymmetric case. More precisely, our way of proceeding will consist, given an admissible superconformal algebra, first to characterize the $M$ generator in the bosonic part, then to determine its stabilizer in the whole superalgebra.
 
 \section{The cases of ${\it SU(2,2/N)}$ with $N \neq 4$ and ${\it pSU(2,2/4)}$}
 
  Any element  $M$ of  ${\it SU(2,2/N)}$  can be written in matrix form, see \cite{DEGKS},  as
\be
(M^*)^{st} \, {\bf L_{4,N}} + {\bf L_{4,N}}\, M = 0  \Longleftrightarrow   M = -{\bf L^{-1}_{4,N}}\, (M^*)^{st} \, {\bf L_{4,N}}  \label{eq:un}
\ee
where $M \in {\it GL(4/N)}$ is given by the following matrix
\be
 M \equiv \left( \begin{array}{cc}
\:  A &   B  \: \\ 
  \:  C    &  D \: \end{array}\right)  \label{eq:84-N}
  \ee
$*$  denotes the complex conjugation and $^{st}$ denotes the supertranspotition
   \be
 M^{st} \equiv \left( \begin{array}{cc}
\:  A^t &   C^t  \: \\ 
  \:  -B ^t    &  D^t\: \end{array}\right)  \label{eq:st}
  \ee
   $A$ is a $4 \times 4$ matrix,   $D$ is a $N \times N$ matrix, $B$ and $C$ are, respectively,  a $4 \times N$ and $N \times 4$ matrix. $A$  and  $B$ are even grading and  $C$ is and  $D$ are odd grading, with the condition 
   \be
   tr A = tr D \label{eq:str}
   \ee
   and  ${\bf L_{4,N}}$ is defined by
   \be
   {\bf L_{4,N}} =   \left( \begin{array}{cc}
\:   {\bf 1_{2,2}}  &  0  \:   \\
  \:  0    &  -i {\bf 1_N}  \: \end{array}\right) 
  \ee
with $ {\bf 1_{2,2}}$ 
\be
 {\bf 1_{2,2}} =   \left( \begin{array}{cc}
\:   {\bf 1_{2}}  &  0  \:   \\
  \:  0    &  - {\bf 1_2}  \: \end{array}\right) 
  \ee
and  ${\bf 1_N}$ being the identity matrix in $N$ dim..
   From eq.(\ref{eq:un})
  \be
  B = i {\bf 1_{2,2}} C^{\dag} \;\;\;\; \Longleftrightarrow  \;\;\;\;  C = i B^{\dag} {\bf 1_{2,2}}  \label{eq:BC}
  \ee 
 and the $A$ and $D$ matrices satisfy
 \be
 A = - {\bf 1_{2,2}} \, A^{\dag} \, {\bf 1_{2,2}} \:\:\:\:\:\:\:\:\:\:\:\:   D = - D^{\dag}  
 \ee
 where $\dag$ stands for the hermitian conjugation.
 The 15 generators in $A$ split into two sets of compact and non compact generators which are, respectively, given by ($j,k = 1,2,3$)
 \be
 \left( \begin{array}{cc}
\:  i \sigma_j &   0  \: \\ 
  \:  0    &  0 \: \end{array}\right) \:\:\:\:\:\:\:\:\:\:\:\:  \left( \begin{array}{cc}
\:  0 &   0  \: \\ 
  \:  0    &   i \sigma_j  \: \end{array}\right) \:\:\:\:\:\:\:\:\:\:\:\: \left( \begin{array}{cc}
\:  i {\bf 1_2} &   0  \: \\ 
  \:  0    &  -i{\bf 1_2} \: \end{array}\right) \label{eq:csu22}
  \ee
  \be
   \left( \begin{array}{cc}
\:  0 &   \sigma_k  \: \\ 
  \:  \sigma_k    &  0 \: \end{array}\right) \:\:\:\:\:\:\:\:\:\:\:\:  \left( \begin{array}{cc}
\:  0 &   i \sigma_k  \: \\ 
  \:  -i \sigma_k    &   0 \  \: \end{array}\right) \:\:\:\:\:\:\:\:\:\:\:\:  \left( \begin{array}{cc}
\: 0  &    i{\bf 1_2}   \: \\ 
  \:   -i{\bf 1_2}     & 0\: \end{array}\right) \:\:\:\:\:\:\:\:\:\:\:\: \left( \begin{array}{cc}
\: 0  &    {\bf 1_2}   \: \\ 
  \:   {\bf 1_2}     & 0\: \end{array}\right) 
  \ee
One recognizes  the maximal compact subalgebra {\it SU(2)} $ \oplus $ {\it SU(2)} $ \oplus  $ {\it  U(1)}
in  eq.(\ref{eq:csu22}), the Lorentz algebra  {\it SO(3,1)} generated by the  rotations, $J_j$ and the boosts $B_k$
\bea
J_j & = &  \left( \begin{array}{cc}
\:  i \sigma_j &   0  \: \\ 
  \:  0    &   i \sigma_j  \: \end{array}\right) \\
   B_k & = &    \left( \begin{array}{cc}
\:  0 &   \sigma_k  \: \\ 
  \:  \sigma_k    &  0 \: \end{array}\right) 
  \eea
 Finally the dilatation $d$, the translations $p_{\mu}$ and the special conformal transformations $c_{\mu}$
 ($\mu = 0,1,2,3$) can be chosen as
  \bea
  d &  = &  \left( \begin{array}{cc}
\: 0  &    {\bf 1_2}   \: \\ 
  \:   {\bf 1_2}     & 0\: \end{array}\right)  \\
 p_j & = & i \left( \begin{array}{cc}
\:   \sigma_j &    \sigma_j  \: \\ 
  \:   -\sigma_j    &   - \sigma_j  \: \end{array}\right) \;\;  \;\;\;\; \;\;  \;\; p_0 = i \left( \begin{array}{cc}
\:    1&     1 \: \\ 
  \:   -1  &   -1 \: \end{array}\right) \\
c_j & = & i \left( \begin{array}{cc}
\:   \sigma_j &   - \sigma_j  \: \\ 
  \:   \sigma_j    &   - \sigma_j  \: \end{array}\right) \;\; \;\; \;\; \; \;\ \; \; \;\; c_0 = i  \left( \begin{array}{cc}
\:    1&     -1 \: \\ 
  \:   1  &   -1 \: \end{array}\right) 
  \eea
  Now,  it is useful to introduce the matrices  $E_{mn}$   with entries  $e_{kl}= \delta_{mk}  \delta_{nl}$, which satisfy, from the property $E_{mn}\, E_{kl} = \delta_{nk} \, E_{ml}$, the commutation  or anti-commutation  relations
 \be
 [E_{mn}, \, E_{kl}]_{\mp} = \delta_{nk} \, E_{ml}  \mp  \delta_{ml} \, E_{kn} 
 \ee 
 Then, the generators of the $D$ part can be chosen as:
 \bea
 && E_{4+p,4+q} -  E_{4+q,4+p}  \;\; \;\; \;\; \;\;  \;\; \;\; p,q = 1, \ldots, N, \; p \neq q \nonumber \\
 &&  i(E_{4+p,4+q} + E_{4+q,4+p}) \nonumber \\
  &&  i(E_{4+r,4+r} -  E_{5+r,5+r}) \;\; \;\; \;\; \;\;  \; r  = 1, \ldots, N-1
  \eea
  There is one last generator in the  ``block-diagonal"  part of $A + D$ that one can choose, in order to ensure the supertrace condition eq.(\ref{eq:str}), as
  \be
  X_N = i \left( N \, \sum_{l=1}^4 \, E_{ll} + 4 \, \sum_{p=1}^N \, E_{4+p,4+p} \right)
  \ee
  In the case $N = 4$, this diagonal generator obviously becomes a multiple of the identity and has then to be eliminated. Considering the quotient of ${\it SU(2,2/4)}$ by the one dimensional ideal generated by $X_4$, one naturally gets ${\it pSU(2,2/4)}$.
  
  A basis for the ``antidiagonal"  $B \oplus C$  can also be commonly obtained in terms of $E_{kl}$ matrices: ($a=1,2; \; p= 1, \ldots, N$)
  \bea
 E_{a,4+p} + iE_{4+p,a}   & \;\; \;\; \;\; \;\; \;\; &  iE_{a,4+p} + E_{4+p,a}   \nonumber \\
 E_{a+2,4+p} - iE_{4+p,a+2}   & \;\; \;\; \;\; \;\; \;\; &  iE_{a+2,4+p} - E_{4+p,a+2}  
    \eea
  In order to identify the elements of the Schr\"odinger algebra in the notations of \cite{BPS}, as explicited in Section 2,  we introduce the following  generators
\bea
J   & =  & \frac{i}{2} \left(\sigma_3  \right) \otimes  \left( \begin{array}{cc}
\: \: 1 &   0  \:\: \\ 
  \: \: 0    &  1 \:\: \end{array}\right) \\
H = \frac{1}{2} \left(p_0 - p_3  \right) & =  & \frac{i}{2} \left(1 + \sigma_3  \right) \otimes  \left( \begin{array}{cc}
 -1 &   -1  \: \\ 
    1    &  1 \: \end{array}\right) \\
  C = \frac{1}{2} \left(c_0 + c_3  \right) & =  & \frac{i}{2} \left(1 + \sigma_3  \right) \otimes  \left( \begin{array}{cc}
\:  1 &   -1  \: \\ 
  \:  1    &  -1 \: \end{array}\right) \\
  D   & =  & \frac{1}{2} \left(1 + \sigma_3  \right) \otimes  \left( \begin{array}{cc}
\: \: 0 &   1  \:\: \\ 
  \: \: 1    &  0 \:\: \end{array}\right) 
 \eea
$J$  is the generator of a $SO(2)$ algebra and   $H, C, D$ span a $SU(1,1)$ algebra 
    \be
    [C, \, H] = - 4 D  \;\;\;\;\; \;\;\;\;\;     [D, \, C] =    2 C \;\;\;\;\; \;\;\;\;\;  [D, \, H] = - 2 H
    \ee
  Let us define 
    \bea
&& K_1  =   \frac{1}{2}  \left( \begin{array}{cc}
 i \sigma_2 &   \sigma_1  \: \\ 
    \sigma_1    &  i \sigma_2 \: \end{array}\right)  
\;\;\;\;\; \;\;  \; \; \;\; \;\;K_2  =   -\frac{1}{2}  \left( \begin{array}{cc}
 i \sigma_1 &   -\sigma_2  \: \\ 
   - \sigma_2    &  i \sigma_1 \: \end{array}\right) \\
 &&  P_1 =  \frac{i}{2} \sigma_1 \otimes \left( \begin{array}{cc}
  1 &   1  \: \\ 
    -1    &  -1 \: \end{array}\right)  \;\;\;\;\; \; \; P_2 =  \frac{i}{2} \sigma_2 \otimes \left( \begin{array}{cc}
  1 &   1  \: \\ 
    -1    &  -1 \: \end{array}\right) 
 \eea   
The $K_a, P_a$ ($a, b = 1,2$) define a $2$-dim Heisenberg algebra ${\it H(2)}$
\be
 [K_a \, P_b] = \delta_{ab} \, M  \;\;\;\;\;\;\;\;\;\;\;\;\;\; [M, \, P_a] = [M, \, K_a] = 0  
 \ee
 where 
 \be
    M = \frac{1}{2} \left(p_0 + p_3  \right)  =   \frac{i}{2} \left(1- \sigma_3  \right) \otimes  \left( \begin{array}{cc}
\:  -1 &   -1  \: \\ 
  \:  1    &  1 \: \end{array}\right) \\  \label{eq:M}
\ee
  We have also the commutations relations  ($a, b = 1,2$)
   \bea
 && [\Lambda, \, P_a] = -2 P_a  \;\; \;\;  \;\;\;\;\; \;\,  [\Lambda, \, K_a] = -2 K_a    \\
 && [J_3, \, P_a] = \vr_{3ab} P_b \;\; \;\;   \;\;\;\;\;  [J_3, \, K_a] = -\vr_{3ab} K_b   
  \eea
  where
 \be
 \Lambda  =   \frac{i}{2} \left(1- \sigma_3  \right) \otimes  \left( \begin{array}{cc}
\:  0 &   1  \: \\ 
  \:  1    &  0 \: \end{array}\right) \label{eq:Lambda}
\ee
The second $2$-dim Heisenberg algebra ${\it H^*(2)}$ is spanned by
  \bea
 && K^*_1  =   \frac{i}{2}  \sigma_1 \otimes \left( \begin{array}{cc}
  1 &   -1  \: \\ 
    1    &  -1 \: \end{array}\right)   \;\;\;\;\; \;\;  K^*_2  =   \frac{i}{2}  \sigma_2 \otimes \left( \begin{array}{cc}
 1 &   -1  \: \\ 
    1   &  -1 \: \end{array}\right) \\
 &&  P^*_1 =  \frac{1}{2} \left( \begin{array}{cc}
 i \sigma_2 &   -\sigma_1  \: \\ 
    -\sigma_1    &  i \sigma_2 \: \end{array}\right)   \;\; \;\;\;\;\; \;  \, P^*_2  =     -\frac{1}{2} \left( \begin{array}{cc}
 i \sigma_1 &   \sigma_2  \: \\ 
    \sigma_2    &  i \sigma_1 \: \end{array}\right)  \\
    &&  M^*  =   \frac{i}{2} \left(1 - \sigma_3  \right) \otimes  \left( \begin{array}{cc}
\:  1 &   -1  \: \\ 
  \:  1    &  -1 \: \end{array}\right) 
 \eea
 \be
 [K^*_a, \, P^*_b ] =   \delta_{ab} M^*  \;\;\;\;\;\;\;\;\;\;\;\;\;\;  [M^*, \, P^*_a] = [M, \, K^*_a] = 0 
 \ee
 Note that the elements $M, M^*$  and $\Lambda$ generate a (second) ${\it SU(1,1)}$ algebra, 
 which commutes with the ${\it SU(1,1)}$  generated by $D, C$ and $H$. Their commutation relations read
\be
 [\Lambda, \, M] = - 2 M \;\;\;\;\;\;\;\;\;\; \;\; [\Lambda, \, M^*] =  2 M^*   \;\;  \;\;\;\;\;\;\;\;\;\; [M, \, M^*] =  -4 \Lambda  \label{eq:MML}
\ee
the dilatation $\Lambda$ acting on ${\it H_2} $ and ${\it H^*_2}$ as ($a,b=1,2$)
   \be
 [\Lambda, \, P_a] = - P_a  \;\;\;\;\;\;\;  [\Lambda, \, P^*_a] =  P^*_a  \;\;\;\;\;\;\;  [\Lambda, \, K_a] = - K_a   \;\;\;\;\;\;\;  [\Lambda, \, K^*_a] =  K^*_a
 \ee
 Finally the commutation relations between {\it H(2)}  and ${\it H^*(2)}$ generators 
 \be
  [K_a \, K^*_b] =   \delta_{ab} C \;\;\;\;\;\;\;\;\;\; \;\; [P_a, \, P^*_b ] =   \delta_{ab} H   \;\;\;\;\;\;\;\;\;\; \;\;
 [P_a, \, K^*_b ] =   -2 \vr_{ab3} J_3 +  \delta_{ab} (D + \Lambda)
 \ee
 provide the {\it SO(2)} $ \oplus $  {\it SU(1,1)} $ \oplus $ {\it SO(1,1)} algebra generated by $J$;  $H,C,D $ and $\Lambda$ respectively. 
 
 Let us emphasize that the {\it SO(2)} $ \oplus  $ {\it SU(1,1)} part is the common bosonic algebra of the two
 Schr\"odinger algebras {\it $ \widetilde{Sch}(2)$}  and {\it $  \widetilde{Sch}^*(2)$} of, respectively, eq.(\ref{eq:sch1}) and eq.(\ref{eq:sch}).
 
 Now, let us turn our attention to the fermionic generators of the {\it SU(2,2/N)} superalgebra. It is straightforward to notice that the $4N$ generators, that we can group into the 4 $N$-vectors of the {\it SU(N)} part: ($p = 1,\ldots,N$)
 \bea
&& Q^1_p =  E_{1,4+p} + i E_{4+p,1}  \;\;\;\;\;\;\;\;\;\; \;\;   Q^2_p = iE_{1,4+p} + E_{4+p,1}
 \nonumber \\
&& S^1_p =  E_{3,4+p}- i E_{4+p,3}  \;\;\;\;\;\;\;\;\;\; \;\;   S^2_p = iE_{3,4+p} - E_{4+p,3}
\eea
commute simultaneously with $M$ and $M^*$. We remark also that each couple ($Q^1_p , Q^2_p$)
and ($S^1_p , S^2_p$) form a doublet under the $J$-rotation. There are $2N$ more generators commuting with $M$ (but not with $M^*$) that we can also consider as 2 N-{\it SU(N)} vectors or 
N  J-{\it SO(2)} doublets: 
\bea
\Xi^1_p & = & \left(E_{4+p.2} +  E_{4+p,4}\right) + i\left(E_{2,4+p}  - E_{4,4+p}\right) \nonumber \\
 \Xi^2_p & = & \left(E_{2,4+p} -  E_{4,4+p}\right) + i\left(E_{4+p,2}  + E_{4+p,4}\right)
 \eea
 and finally $2N$ generators commuting with $M^*$ (but not with $M$) again classified in the representation (2,N) of  {\it SO(2)} $ \oplus  $ {\it SU(N)}.
 \bea
\Xi^{(*)1}_p & = & \left(E_{4+p.2} -  E_{4+p,4}\right) + i\left(E_{2,4+p}  + E_{4,4+p}\right) \nonumber \\
 \Xi^{(*)2}_p & = & \left(E_{2,4+p} +  E_{4,4+p}\right) + i\left(E_{4+p,2}  - E_{4+p,4}\right)
 \eea
 The anti-commutation relations between  the $Q^a_p$  and $S^b_q$ ($a,b=1,2; \; p,q=1,\ldots,N$)
can be summarised as follows
  \bea
 \{Q^1_p ,\, Q^2_q\} & = &-  \{S^1_p,  \, S^2_q\}  =  - (E_{4+p.4+q} - E_{4+q,4+p}) \nonumber \\
 \{Q^1_p, \, S^1_q\} & = & \{Q^2_p,  \, S^2_q\}  =  \delta_{pq} \frac{H+C}{2}   \nonumber \\
  \{Q^1_p, \, S^2_q\} & = &  - \{Q^2_p,  \, S^1_q\}  =  - \delta_{pq} D   \nonumber \\
 \{Q^1_p, \, Q^1_q\} & = &  \{Q^2_p,  \, Q^2_q\}  =  i (E_{4+p.4+q} + E_{4+q,4+p}) \;\; \;\;   (p \neq q) \nonumber \\
  \{Q^1_p, \, Q^1_p\} &  = &   \{Q^2_p,  \, Q^2_p\}  =  J - \frac{H-C}{2} + \frac{1}{2N}  \, X_N   +  Z_{N;p}  \nonumber \\
  \{S^1_p, \, S^1_q\}  & =  & \{S^2_p,  \, S^2_q\}  =  -   \{Q^1_p,\, Q^1_q\} \;\;\;\;   (p \neq q) \nonumber \\
  \{S^1_p, \, S^1_p\} & = &  \{S^2_p,  \, S^2_p\}  =   - \{Q^1_p ,\, Q^1_p\} - (H-C)   
 \eea
 where
 \bea
 Z_{N;p}  & = &  - \frac{2i}{N}(E_{55} - E_{66}) - \frac{4i}{N}(E_{66} - E_{77}) - \ldots - \frac{2(p-1)i}{N}(E_{3+p.3+p} - E_{4+p,4+p})  \nonumber \\
  &  &  + \frac{2(N-p)i}{N}(E_{4+p.4+p} - E_{5+p,5+p}) + \frac{2(N-p-1)i}{N}(E_{5+p,5+p} - E_{6+p,6+p}) 
  \nonumber \\
  & & + \ldots + \frac{2i}{N}(E_{3+N.3+N} - E_{4+N,4+N}) 
  \eea
Note that, for $N=4$, $X_4$ disappears on the r.h.s. of the above anti-commutation relations when considering {\it pSU(2,2/4)}.  We notice that all these anti-commutation relations close into the elements of $H,C,D$ of {\it SU(1,1)}, the elements of {\it SU(N)} and the  {\it SO(2)} generator $\left(J +\frac{X_N}{2N}\right)$ which commutes with  {\it SU(1,1)} as well as with {\it SU(N)}. Adding to this reductive algebra 
{\it SU(1,1)} $ \oplus $ {\it SU(N)}  $ \oplus  $  {\it SO(2)} the $4N$ elements $\, Q^a_p, S^a_p$ with $a=1,2; p=1,\ldots,N$, one obtains a realization  of  the {\it SU(1,1/N)} superalgebra.

At this point, let us consider more precisely the action of  $J $ and $X_N$ on the fermionic sector; one gets
\be
 [J, \, Q^a_p]  =  \frac{1}{2} \, \vr^a_b \, Q^b_p \;\;\;\; \;\;\;\; \;\;\;\;
[J, \, S^a_p]  =  \frac{1}{2} \, \vr^a_b \, S^b_p  \;\;\;\;\;\;\;\; \;\;\;\;
 [J, \, \Xi^{(*)a}_p] =  \frac{1}{2} \, \vr^a_b \, \Xi^{(*)b}_p
\ee
$\vr^a_b$ being the $2 \times 2$ anti-symmetric tensor with $\vr^1_2 = -\vr^2_1=1$, and
\be
 [X_N, \, Q^a_p]  =  (N-4) \, \vr^a_b \, Q^b_p \;\;\;\; \;\;\;\; 
[X_N, \, S^a_p]  =  (N-4) \, \vr^a_b \, S^b_p  \;\;\;\;\;\;\;\; 
 [X_N, \, \Xi^{(*)a}_p] =  -(N-4)  \, \vr^a_b \, \Xi^{(*)b}_p
\ee
It follows that the rotation generator $J$ in the bosonic  Schr\"odinger  algebra appears, for $N \neq 4$, as a combination of the two compact generators  $\left(J +\frac{X_N}{2N}\right)$ in {\it SU(1,1/N)} and
 $\left(J -\frac{X_N}{2(N-4)}\right)$, this last element commuting with {\it SU(1,1/N)}, with which it therefore constitutes the direct sum  $ {\it SO(2)} \oplus {\it SU(1,1/N)}$.

Now, adding to the Heisenberg algebra  {\it H(2)}, generated by $P_a, K_a$ ($a=1,2$) and $M$, the fermionic generators $\Xi^a_p$ ($p=1,\ldots,N$) one obtains the super-Heisenberg algebra {\it SH(2/N)}.
In the same way, the elements $P^*_a, K^*_a,\Xi^{(*)a}_p$ and $M^*$ generate the super-Heisenberg algebra ${\it SH^*(2/N)}$. One easily verifies the action of  {\it SU(1,1/N)} on {\it SH(2/N)} and, similarly, on ${\it SH^*(2/N)}$.

Let us summarise our results:
\begin{itemize}
\item The  super Schr\"odinger algebra in {\it SU(2,2/N)} with $N \neq 4$,  extension of the Schr\"odinger algebra in {\it SU(2,2)} and stabilizer of the bosonic generator $M$,  see eq.(\ref{eq:M}), shows up as the semi-direct sum of the above defined algebra ${\it SO(2)} \oplus {\it SU(1,1/N)}$  on the super-Heisenberg algebra   {\it SH(2/N)}, that is: 
\be
[{\it SO(2)} \oplus {\it SU(1,1/N)}] \rhd {\it SH(2/N)}
\ee
Another copy, conjugate in {\it SU(2,2/N)} of super Schr\"odinger algebra 
is provided by:
\be
[{\it SO(2)} \oplus {\it SU(1,1/N)}] \rhd {\it SH^*(2/N)}
\ee
and a ``schematic" decomposition of the {\it SU(2,2/N)}, $N \neq 4$, superalgebra is the following:
\be
{\it SU(2,2/N)} =_{v.s.}  {\it SH(2/N)} \lhd [{\it SO(2)} \oplus{\it SU(1,1/N)}) \oplus  {\it SO(1,1)}] \rhd {\it SH^*(2/N)}
\ee
the last  {\it SO(1,1)} in the above expressions being generated by $\Lambda$, defined in 
eq.(\ref{eq:Lambda}). We remind that $\Lambda, M$ and $M^*$ close into a {\it SU(1,1)}, see  eq.(\ref{eq:MML}).
\item In the case $N=4$. the relevant super-conformal algebra is {\it pSU(2,2/4)}  and the above extra {\it SO(2)}  is no more present in the super Schr\"odinger sector, which then reduces to
\be
{\it SU(1,1/4)} \rhd {\it SH(2/4)}
\ee
and also to:
\be
{\it SU(1,1/4)} \rhd {\it SH^*(2/4)}
\ee
providing the vector space decomposition:
\be
{\it pSU(2,2/4)} =_{v.s.}  {\it SH(2/4)} \lhd [{\it SU(1,1/4)} \oplus  {\it SO(1,1)}] \rhd {\it SH^*(2/4)}
\ee
\end{itemize}
  
 \section{The case of {\it Osp(N/4,R)}}

Any element  $M$ of {\it Osp(N/4,R)} can be written in matrix representation on {\bf R} as
\be
 M \equiv \left( \begin{array}{cc}
\:  A &   B  \: \\ 
  \:  C    &  D \: \end{array}\right)  \label{eq:84}
  \ee
  with $A$ and $D$ being respectively  a $N \times N$ and a $4 \times 4$ matrix (even grading part) and $B$ and $C$, respectively,  a $N \times 4$ and $4 \times N$ matrix (odd grading part), with the following condition, see \cite{DEGKS}
\be
M^{st} \, {\bf K}  + {\bf K}  \, M = 0  \Longleftrightarrow   M = - {\bf K^{-1}}  \, (M)^{st} \, {\bf K}   \label{eq:ort}
\ee
$^{st}$ denotes the supertransposition defined in eq.(\ref{eq:st})
  and $K$  stands for the matrix
  \be
{\bf K} =  \left( \begin{array}{cc}
 \:  {\bf 1_N} &   0  \: \\ 
  \:  0    & {\bf J_4} \: \end{array}\right) 
  \ee 
 with ${\bf J_{4}}$  
 \be
{\bf J_{4}} =   \left( \begin{array}{cc}
\:  0 &   {\bf 1_2}   \: \\ 
  \: - {\bf 1_2}    &  0 \: \end{array}\right) 
  \ee
One easily gets:
 \be
 A = - A^t \;\;\;\;\;  \; \; B = - C^t {\bf J_4} \;\;\;\;\;  \; \; C = - {\bf J_4} B^t \;\;\;\;\;\; \; D = {\bf J_4} D^t {\bf J_4}  \label{eq:rf1}
 \ee
from which an explicit basis for {\it Osp(N/4,R)} can be derived with the $A$ part - or {\it SO(N)} algebra - generated by
\be
A_{ij} = E_{ij}  - E_{ji} = = - A_{ji} \;\;\;\;\; (i,j = 1,2, \ldots, N) 
\ee
the $D$ part - or  {\it Sp(4,R)} $ \cong $ {\it SO(3,2)}  algebra - generated by
\bea
&& E_{N+1,N+1}  - E_{N+3,N+3} \;\;\;\;\;   \;\;\;\;\; \;\;\;\;\; E_{N+2,N+2}  - E_{N+4,N+4} \nonumber \\
&& E_{N+1,N+2}  - E_{N+4,N+3} \;\;\;\;\;  \;\;\;\;\;   \;\;\;\;\;  E_{N+2,N+1}  - E_{N+3,N+4}  \nonumber \\
&&  E_{N+1,N+3}, \;\;\;\;\;\; E_{N+2,N+4}, \;\;\;\;\; \;\;\;\;\;  E_{N+3,N+1},  \;\;\;\;\;\, E_{N+4,N+2} \nonumber \\
&& E_{N+3,N+2}  + E_{N+4,N+1} \;\;\;\;\; \;\;\;\;\;  \;\;\;\;\;  E_{N+1,N+4}  + E_{N+2,N+3}
\eea
and finally the  $B$ and $C$ or fermionic part spanned  by ($i= 1,2, \ldots, N$) :
\bea
&& E_{i,N+1}  + E_{N+3,i} \;\;\;\;\; \;\;\;\;\;  \;\;\;\;\;  E_{i,N+2}  + E_{N+4,i} \nonumber \\
&& E_{i,N+3}  - E_{N+1,i} \;\;\;\;\; \;\;\;\;\; \;\;\;\;\;   E_{i,N+4}  - E_{N+2,i}
\eea  
Since the algebraic results which follow are similar whatever the value of the positive integer $N$, in the following we will choose $N=8$, which is the relevant value in the present $AdS \times S^7$ M-theory.
First focusing our attention to the {\it Sp(4,R)} $ \cong $ {\it SO(3,2)} algebra, we can recognize the subalgebras
\be
{\it SO(2,2)}  \cong  {\it SO(2,1)}  \oplus {\it SO(2,1)}
\ee
with basis
\be
\{E_{9,11}, E_{11,9},E_{9,9} - E_{11,11} \} \oplus \{E_{10,12},E_{12,10}, E_{10,10} - E_{12,12} \} \label{eq:so21}
\ee
We shall take the first three generators to generate the $H, C, D$ elements and the other three to get the $M, M^*$ and $\Lambda$ elements.
Then the two $d=1$ Heisenberg algebras show up with {\it H(1)} generated by:
\be
P = E_{9,12} + E_{10,11}\;\;\;\;\;  \;\;\;\;\; K = E_{10,9}- E_{11,12} \;\;\;\;\;  \;\;\;\;\; M = 2 \, E_{10,12} 
\ee
and ${\it H^*(1)}$ generated by:
\be
P^* = E_{9,10} - E_{12,11}\;\;\;\;\; \;\;\;\;\; K^* = E_{12,9} + E_{11,10}\;\;\;\;\;  \;\;\;\;\; M^* = 2 \, E_{12,10} 
\ee
Considering the fermionic part, one can check that each of the $16$ generators ($ i =1,2,\ldots,8$)
\be
  Q_{i}  =  E_{i,9} + E_{11,i} \;\;\;\;\;  \;\;\;\;\; \;\;\;\;\;  S_{i}  =  E_{i,11} - E_{9,i} 
\ee
 commutes both with  $M$ and $M^*$. Moreover, the $8$ generators
 \be
\Xi_{i}  =  E_{i,12} - E_{10,i} 
\ee
 commute with  $M$, but not with  $M^*$, while the $8$ generators
 \be
\Xi^*_{i}  =  E_{i,10} + E_{12,i} 
\ee
 commute with  $M^*$, but not with  $M$.
 
 It is a simple exercise to verify some more commutation relations and to deduce that the Heisenberg algebra
{\it H(1)}  can be  extended to {\it SH(1/8)} by adding to it the elements $\Xi_{i}$. In the same way  ${\it H^*(1)}$ can be extended to ${\it SH^*(1/8)}$ by adding the $\Xi^*_{i}$ elements.

 As it could be expected, the anti-commutation relations between the $Q_{i}$ and  $S_{j}$  ($ i ,j =1,2,\ldots,8$) provide the generators of the  {\it SO(8)} as well as of  {\it Sp(2,R)} algebras generated by $H, C, D$:
\bea
\{Q_{i}, \, S_{j} \}&  = & \delta_{ij}  \, (E_{11,11} - E_{9,9}) - A_{ij}  \nonumber \\
\{Q_{i}, \, Q_{j} \}&  = &  2 \, \delta_{ij}  \, E_{11,9} \nonumber \\
\{S_{i}, \, S_{j} \}  & = & - 2  \, \delta_{ij} \,  E_{9,11} 
\eea
Moreover, one can prove that  the $Q_{i}'s$ and  $S_{i}'s$  form the fermionic part of the {\it Osp(8/2,R)}  superalgebra, the bosonic part being {\it SO(8)} $ \oplus $ {\it Sp(2,R)}. The super Schr\"odinger algebra
in {\it Osp(8/4,R)}  can finally be seen as the semi-direct sum of {\it Osp(8/2,R)}  acting on {\it SH(1/8)} 
\be
{\it Osp(8/2,R)}  \rhd {\it SH(1/8)} 
\ee
or on ${\it SH^*(1/8)}$
\be
{\it Osp(8/2,R)}  \rhd {\it SH^*(1/8)} 
\ee
 leading to the  ``schematic decomposition"  of  {\it Osp(8/4,R)}
 \be
 {\it Osp(8/4,R)} =_{v.s.} {\it SH(1/8)}  \lhd [{\it Osp(8/2,R)}  \oplus {\it SO(1,1)}] \rhd {\it SH^*(1/8)} 
 \ee
 with 
 {\it SO(1,1)} generated by 
 \be
 \Lambda = E_{10,10} - E_{12,12}
 \ee
 This decomposition holds for any non negative value of the integer $N$ 
  \be
 {\it Osp(N/4,R)} =_{v.s.} {\it SH(1/N)}  \lhd [{\it Osp(N/2,R)}  \oplus {\it SO(1,1)}] \rhd {\it SH^*(1/N)} 
 \ee
 in which the super Schr\"odinger algebra reads: 
 \be
 {\it Osp(N/2,R)} \rhd {\it SH(2/N)} 
 \ee

\section{The case of {\it Osp(6,2/2N)}}

This case presents the peculiarity that the $R$-symmetry is described by a (compact form of) symplectic algebra. As we will see, this leads to results slightly different from the ones obtained when the $R$-symmetry is a unitary or orthogonal algebra. Now, using again the definition of \cite{DEGKS}, any element of {\it Osp(6,2/2N)} can be written as a $(8 + 2N) \times (8 + 2N)$ matrix of {\it Osp(8/2N)}, see
eq.(\ref{eq:ort}), with the extra condition:
\be
 M = - {\bf \tilde{K}}^{-1} \, (M^*)^{st} \, {\bf \tilde{K}}  \label{eq:ncf}
\ee
where  ${\bf \tilde{K}}$ is the matrix
  \be
{\bf \tilde{K}} =  \left( \begin{array}{cc}
 \:  {\bf J_8} &   0  \: \\ 
  \:  0    & {\bf 1_{2N}} \: \end{array}\right) 
  \ee 
the other symbols in eq.(\ref{eq:ncf}) being already defined in Section 4 and 5.

Let us, for the time being, develop our computation for the case {\it Osp(6,2/4)}. We will come back to the general case  {\it Osp(6,2/2N)} at the end of this Section. Then, any element $M$ of the considered superalgebra can be represented by the $12 \times 12$ matrix:
\be
 M =  \left( \begin{array}{cc}
\:  \hat{A} &   B  \: \\ 
  \:  C  &  D \: \end{array}\right)  \label{eq:osp624N}
  \ee
 Eq.(\ref{eq:ncf}) implies for the $8 \times 8$ matrix $\hat{A}$ and for the $4 \times 4$ matrix $D$   and for the matrices $B$ and $C$ the relations
 \be
 \hat{A} = {\bf J_8} \hat{A}^{*t} {\bf J_8} = -\hat{A}^t, \;\;\;\;\;  D =  {\bf J_8} D^t {\bf J_4} = - D^{*t},  \;\;\;\;\;
 B = {\bf J_8} B^* {\bf J_4},\;\;\;\;\; C = - {\bf J_4}  B^t \  \label{eq:rf2}
  \ee
 In this framework, the {\it SO(6,2)} algebra, corresponding to the matrices $\hat{A}$ in eq.(\ref{eq:osp624N}), can be convemiently described in the following basis 
 ($j= 1,2,\ldots,6; \, k= 1,2,\ldots,10$)
 \be
\hat{A}_j = \left( \begin{array}[tl]{cc}
\:  A_j &   0 \: \\ 
  \:  0   &  A_j  \: \end{array}\right)
  \;\;\;\;\; \;\; \;\;\;\;\;
\hat{S}_k = \left( \begin{array}[tr]{cc}
\:  0  &   S_k \: \\ 
  \:  -S_k  &  0  \: \end{array}\right)
  \ee
  for the compact part and 
  \be
\hat{B}_j = \left( \begin{array}{cc}
\: i A_j &   0 \: \\ 
  \:  0   &  -iA_j  \: \end{array}\right)
 \;\;\;\;\;  \;\;\;\;\; 
\hat{C}_j = \left( \begin{array}{cc}
\:  0  &   i A_j \: \\ 
  \:  i A_j  &  0  \: \end{array}\right)
  \ee
for the non compact part, 
$A$ being $4 \times 4$ real antisymmetric matrices, defined as follows:
\bea
 A_1 & = &   (E_{12} - E_{21}) + (E_{34} - E_{43}) \nonumber \\
  A_2 & = & -(E_{23} -  E_{32})  -  (E_{14}  - E_{41}) \nonumber \\
   A_3 & = &  -(E_{13} - E_{31}) + (E_{24}  -  E_{42})  \nonumber \\
   A_4 & = &   (E_{12} - E_{21}) -  (E_{34} - E_{43}) \nonumber \\
   A_5  & = &    (E_{23} -  E_{32})   -  (E_{14}  - E_{41})\nonumber \\
   A_6 & = &  -(E_{13} - E_{31})-  (E_{24}  -  E_{42})  
   \eea
   with $\{A_1, A_2, A_3\} $ and $ \{A_4, A_5, A_6\}$  generating two commuting {\it SO(3)}. 
  $S$ are $4 \times 4$ real a symmetric matrices, defined as follows
\bea
   S_1 & = &   (E_{23} + E_{32}) - (E_{14} + E_{41})   \nonumber \\
 S_2 & = & (E_{12} +  E_{21} ) +  (E_{34}  + E_{43})  \nonumber  \\    
   S_3 & = &  (E_{11} -  E_{22}+ E_{33}  -  E_{44}) \nonumber \\
  S_4 & = &   (E_{11} - E_{33} + E_{22} - E_{44})   \nonumber \\
 S_5 & = & (E_{23} + E_{32}) + ( E_{14}  + E_{41})  \nonumber  \\    
   S_6 & = &  (E_{13} + E_{31}) - (E_{24}  +  E_{42}) \nonumber \\
   S_7 & = &   (E_{13} + E_{31} + E_{24} + E_{42})   \nonumber \\
 S_8 & = & (E_{11}-  E_{22}  -  E_{33}  + E_{44})  \nonumber  \\    
   S_9 & = &  (E_{12} + E_{21}) - (E_{34}  +  E_{43}) \nonumber  \\ 
   S_{10} & = &   (E_{11} + E_{22} + E_{33}+  E_{44})   
   \eea
   Since {\it SO(6,2)} stands fo the conformal algebra in $5+1$ dimensions, i.e. $5$ space and $1$ time dimensions, the Schr\"odinger counterpart we are looking for will act on a $(4+1)$-dimensional space. Considering the embedding:
   \be
{\it SO(6,2)}  \supset {\it SO(4)} \oplus {\it SO(2,2)}    \cong {\it SO(4)} \oplus {\it SO(2,1)}  \oplus   {\it SO(2,1)} 
\ee
the  ``rotation" algebra {\it SO(4)}, acting on the translation $P_a$ and momenta $K_a$ ($a= 1,2,3,4$), can be chosen as 
\be
{\it SO(4)} \equiv \{\hat{A}_1, \hat{A}_2, \hat{A}_3,  \hat{S}_1,  \hat{S}_2, \hat{S}_3\}  \label{eq:so4}
\ee
The two commuting  {\it SO(2,1)}  algebras can be chosen as
\be
\{\hat{A}_6  -  \hat{S}_{10},   \hat{B}_4  +  \hat{C}_5,  \hat{B}_5  -  \hat{C}_4\} \;\;\;\;  \oplus  \;\;\;\;   \{\hat{A}_6  +  \hat{S}_{10},   \hat{B}_4  -  \hat{C}_5, \hat{B}_5  +  \hat{C}_4\} 
\ee
The first  {\it SO(2,1)}, which we will denote  $ {\it SO(2,1)}_c,$  will be chosen to contain the $H, C, D$ generators.
Due to the commuting relations
\bea
[\hat{A}_6  -  \hat{S}_{10},\, \hat{B}_4  +  \hat{C}_5] = - 4 (\hat{B}_5  -  \hat{C}_4) \;\;\;\;\; && \;\;\;\;\; 
[\hat{A}_6  -  \hat{S}_{10},\, \hat{B}_5  -  \hat{C}_4] = 4 ( \hat{B}_4  +  \hat{C}_5)  \nonumber \\
\mbox{}  [\hat{B}_4  +  \hat{C}_5, \, \hat{B}_5  -  \hat{C}_4] & =&  - 4 (\hat{A}_6  -  \hat{S}_{10})
\eea
one can make the identification
\be
D = \frac{1}{2} \, (\hat{A}_6  -  \hat{S}_{10}) \;\; \;\;\;\;\; \;\; H= \frac{1}{2\sqrt{2}} \, ( \hat{B}_4  +  \hat{C}_5)\;\ \;\;\;\;\;  \;\; C = \frac{1}{2\sqrt{2}} \, (\hat{B}_5  -  \hat{C}_4) \label{eq:o21-1}
\ee
Then,  the second  {\it SO(2,1)}   will contain the generators $M, M^*, \Lambda$ and we will choose
\bea
&& M = (\hat{A}_6  + \hat{S}_{10}) +  (\hat{B}_5  +  \hat{C}_4) \label{eq:M62} \\
&& M^* = (\hat{A}_6  + \hat{S}_{10}) - (\hat{B}_5  +  \hat{C}_4) \label{eq:M*62} \\
&& \Lambda = - 4(\hat{B}_4 -  \hat{C}_5)
\eea
As a final step we have to determine the $4$ translations $P_{a}$ and the $4$ momenta  $K_{a}$
($a =1,2,3,4$). From what we already know, the space translation  $p_{5}$  and time translation $p_{0}$ now appear, in the conformal algebra, as 
\be
p_5 = \hat{A}_6  +  \hat{C}_{4} \;\ \;\;\;\;\;\;\;\;\;\;   p_0 =  \hat{S}_{10} +  \hat{B}_5  
\ee
By action on $p_5$ of the {\it SO(5)} rotation algebra, constructed from the above defined {\it SO(4)}, 
eq.(\ref{eq:so4}),  by addition the $4$ generators $\{\hat{A}_6,  \hat{S}_4,  \hat{S}_5,  \hat{S}_6\}$, one can generate the $p_{a} \equiv P_{a}$ translations and, finally, the momenta $K_{a}$, which have to satisfy the condition
\be
[K_{a},\, P_{b}] = \delta_{ab} \, M \;\;\;\;\;\;\;\;\;\;  (a,b =1,2,3,4)
\ee
We obtain
\bea
P_{1} & =  &- \frac{1}{\sqrt{2}} \,(\hat{S}_7  -  \hat{B}_1) \;\;\;\;\;\;\;\;\;\;\;\; K_{1} = - \frac{1}{\sqrt{2}} \,(\hat{S}_4  +  \hat{C}_1) \nonumber \\
P_{2} & =  & \frac{1}{\sqrt{2}} \, (\hat{S}_8 +  \hat{B}_2)  \;\;\;\;\;\;\;\;\;\;\;\; \;\; \; K_{2} = - \frac{1}{\sqrt{2}} \,(\hat{S}_6  +  \hat{C}_2) \nonumber \\
P_{3} & =  &- \frac{1}{\sqrt{2}} \, (\hat{S}_9  -  \hat{B}_3) \;\; \;\;\;\;\;\;\;\;\;\; K_{3} =  \frac{1}{\sqrt{2}} \,(\hat{S}_5  -  \hat{C}_3) \nonumber \\
P_{4} & =  & \frac{1}{\sqrt{2}} \, (\hat{A}_4  -  \hat{C}_6) \;\; \;\;\;\;\;\;\;\;\;\; \;\; \;K_{4} =  \frac{1}{\sqrt{2}} \, (\hat{A}_5 -  \hat{B}_6) \label{eq:pk}
\eea
The second Heisenberg algebra ${\it H^*(4)}$ is easily obtained  by first operating a change of sign in front of the $\hat{B}$ and $\hat{C}$ (that is replacing $i$ by $-i$ in the non compact generators), appearing in the above combination eq.(\ref{eq:pk}) and  then denoting $K^*_{a}$  the so transformed $P_{a}$ and $P^*_{a}$ the so transformed $-K_{a}$.  More precisely we have 
\bea
P^*_{1} & =  & \frac{1}{\sqrt{2}} \,(\hat{S}_4  -  \hat{C}_1) \;\; \;\;\;\;\;\;\;\;\;\; \;\; \;\; K^*_{1} = - \frac{1}{\sqrt{2}} \,(\hat{S}_7  +  \hat{B}_1) \nonumber \\
P^*_{2} & =  & \frac{1}{\sqrt{2}} \, (\hat{S}_6 -  \hat{C}_2) \;\; \;\;\;\;\;\;\;\;\;\; \;\; \;  \;K^*_{2} =  \frac{1}{\sqrt{2}} \,(\hat{S}_8  -  \hat{B}_2) \nonumber \\
P^*_{3} & =  &- \frac{1}{\sqrt{2}} \, (\hat{S}_5  +  \hat{C}_3) \;\;  \;\;\;\;\;\;\;\;\;\; \; \; K^*_{3} =  - \frac{1}{\sqrt{2}} \,(\hat{S}_9  +  \hat{B}_3) \nonumber \\
P^*_{4} & =  & -\frac{1}{\sqrt{2}} \, (\hat{A}_5  +  \hat{B}_6) \;\;  \;\;\;\;\;\;\;\;\;\; \,K^*_{4} =  \frac{1}{\sqrt{2}} \, (\hat{A}_4 +  \hat{C}_6) \label{eq:pk*}
\eea
As expected, one gets 
\be
[K^*_{a}, \, P^*_{b}] = \delta_{ab} \, M^*  
\ee
Let us close our study of the bosonic part by briefly studying  the algebra {\it Sp(4)}.  A natural basis   for the $4 \times 4$ matrix $D$, satisfying eq.(\ref{eq:rf2}), is provided by the $10$ generators:
\bea
  &  &   i(E_{11} - E_{33}),\;\;\;\;  \;\;\;\; i(E_{22} - E_{44}),    \;\;\;\;  \;\;\;\;  i(E_{13} +  E_{31}),  \;\;\;\;  \;\;\;\;  i(E_{24}  + E_{42})  \nonumber  \\    
    &  &  i(E_{12} + E_{21} -  E_{34}  -  E_{43}),   \;\;\;\;  \;\;\;\; \;\;\;\;\; \;\;\;\;
   i(E_{14} +  E_{41} +  E_{23} + E_{32})    \nonumber  \\    
    &  &  (E_{13} -  E_{31}),\;\;\;\;  \;\;\;\;  \;\;\;\;\;  \;\;\;\; \;\;\;\;\; \;\;\;\;\;  \;\;\;\;\; (E_{12} - E_{21} + E_{34} -  E_{43})   \nonumber  \\ 
   &&  (E_{24}  -  E_{42}),   \;\;\;\;  \;\;\;\; \;\;\;\;\; \;\;\;\;  \;\;\;\;\; \;\;\;\;\;  \;\;\;\;\;  (E_{14} +  E_{23}  -  E_{32}  - E_{41})  \label{eq:spa4}
    \eea
Now, we turn our attention to the fermionic part. A basis for the $32$ fermionic generators, satisfying eq.(\ref{eq:rf2}), can be choosen as: ($i,j = 1,2,3,4$)
  \bea 
  F_{i,9} \; &= & E_{i,9} +  E_{9,4+i} - E_{4+i,11} + E_{11,i} \;\;\;\;\; \;\;  \;  \; F_{i,10} = E_{i,10} +  E_{10,4+i} - E_{4+i,12} + E_{12,i} \nonumber \\
   F_{i,11} & = & E_{i,11} +  E_{11,4+i} +  E_{4+i,9} - E_{9,i} \;\;\;\;\; \;\;  \;  \; F_{i,12} = E_{i,12} +  E_{12,4+i} + E_{4+i,10} - E_{110,i} \label{eq:F} 
   \eea
   and
   \bea
   G_{j,9}  \;  & = & i(E_{j,9} -  E_{9,4+j} + E_{4+j,11} + E_{11,j}) \;\;\;\;\;  G_{j,10} = i(E_{j,10} -  E_{10,4+j} + E_{4+j,12} + E_{12,j}) \nonumber \\
   G_{j,11} & = & i(E_{j,11} -  E_{11,4+j} -  E_{4+j,9} - E_{9,j}) \;\;\;\;\; G_{j,12} = i(E_{j,12} -  E_{12,4+j} - E_{4+j,10} - E_{110,j}) \label{eq:G}
   \eea
   The anti-commutation relations of the above defined fermionic generators are reported in Appendix.
It is straightforward, but a tedious exercise, to determine the fermionic combinations which commute with $M$ and those which commute $M^*$. It appears that $16$ fermions commute with $M$ and $M^*$ simultaneously; they can be choosen as:
\bea
Q_1  & = & F_{1,9} - F_{3,11},   \;\;\;\;\; \;\;  \; \;\;\;\;\;  Q_2  = F_{2,9} - F_{4,11}   \nonumber \\
Q_3 & = & F_{3,9} + F_{1,11}, \;\;\;\;\; \;\;  \;  \;\;\;\;\;  Q_4 =  F_{4,9} + F_{2,11}  \nonumber \\
Q_5 & = & F_{1,10} - F_{3,12}, \;\;\;\;\; \;\; \;\;\;\;\;  Q_6 =  F_{2,10} - F_{4,12}  \nonumber \\
Q_7  & = & F_{3,10} + F_{1,12}, \;\;\;\;\; \;\; \;\;\;\;\;  Q_8 =  F_{4,10} + F_{2,12} \\
S_1 & = &  G_{1,9} + G_{3,11}, \;\;\;\;\; \;\;  \;\;\;\;\;  \; S_2 = G_{2,9} + G_{4,11} \nonumber \\
S_3  & = & G_{3,9} - G_{1,11}, \;\;\;\;\; \;\; \;\;\;\;\; \; S_4 = G_{4,9} - G_{2,11}  \nonumber \\
S_5 & = & G_{1,10} + G_{3,12}, \;\;\;\;\; \;\;  \;\;\;\;\;  S_6 = G_{2,10} + G_{4,12} \nonumber \\
S_7 & = & G_{3,10} - G_{1,12}, \;\;\;\;\; \;\; \;\;\;\;\;  S_8 = G_{4,10} - G_{2,12}
\eea
One can recognize the representations ($1/2, 1/2,  {\bf 4}$) of the algebra ${\it SO(2,1)}_c \oplus {\it SO(3)} \oplus {\it Sp(4)}$  algebra, ${\it SO(2,1)}_c$  being the   ``conformal" one given by eq.(\ref{eq:o21-1}), the {\it SO(3)}  is generated by $\{\hat{A}_1  + \hat{S}_1, \hat{A}_2 -  \hat{S}_2, \hat{A}_3 - \hat{S}_3\}$ and we denote it as  ${\it SO(3)}_{+}$, and {\it Sp(4)} is  the $R$-symmetry of our problem given by eq.(\ref{eq:spa4}). Performing all the anti-commutation relations among the $Q_{\mu}$ and the $S_{\mu}$ ($\mu = 1,2,\ldots,8$) one gets back  the whole above defined semi-simple algebra, so proving that we have a realization of the ${\it Osp(4^*/4)}$ superalgebra. It is important to remark that the ``rotation"  {\it SO(4)} algebra defined by eq.(\ref{eq:so4}) is the semi-direct sum of  the above defined ${\it SO(3)}_{+}$ and of a second {\it SO(3)}, denoted ${\it SO(3)}_{-}$, generated by  $\{\hat{A}_1 - \hat{S}_1, \hat{A}_2 +  \hat{S}_2, \hat{A}_3 + \hat{S}_3\}$. One can check that   ${\it SO(3)}_{-}$, does not act on the $Q_{\mu}$ and the $S_{\mu}$, but effectively acts on the $\Xi_{\mu}$ and the $\Xi^*_{\mu}$  which will respectively be added to the
{\it H(4)}  and ${\it H^*(4)}$ Heisenberg algebra to constitute the {\it SH(4)}  and ${\it SH^*(4)}$ super-Heisenberg ones, and which can be defined as follows:
\bea
\Xi_1 & = & (F_{1,9} + F_{3,11}) - (G_{2,9} - G_{4,11})   \;\;\;\;\;\;\;\;\;\; \;\;\; \Xi_2   =  (F_{2,9} + F_{4,11}) + (G_{1,9} - G_{3,11})  \nonumber \\
\Xi_3 & = & (F_{3,9} - F_{1,11}) + (G_{4,9} + G_{2,11})   \;\;\;\;\;\;\;\;\;\; \;\;\;  \Xi_4   =  (F_{4,9} - F_{2,11}) - (G_{3,9} + G_{1,11})  \nonumber \\
\Xi_5 & = & (F_{1,10} + F_{3,12}) - (G_{2,10} - G_{4,12})   \;\;\;\;\;\;\;\;\;\;  \Xi_6   =  (F_{2,10} + F_{4,12}) + (G_{1,10} - G_{3,12})  \nonumber \\
\Xi_7 & = & (F_{3,10} - F_{1,12}) + (G_{4,10} + G_{2,12})   \;\;\;\;\;\;\;\;\;\;   \Xi_8   =  (F_{4,10} - F_{2,12}) - (G_{3,10} + G_{1,12}) \label{eq:Xi}
\eea
and
\bea
\Xi^*_1 & = & (F_{1,9} + F_{3,11}) + (G_{2,9} - G_{4,11})   \;\;\;\;\;\;\;\;\;\; \;\;\; \Xi^*_2   =  (F_{2,9} + F_{4,11}) -  (G_{1,9} - G_{3,11})  \nonumber \\
\Xi^*_3 & = & (F_{3,9} - F_{1,11}) - (G_{4,9} + G_{2,11})   \;\;\;\;\;\;\;\;\;\; \;\;\;  \Xi^*_4   =  (F_{4,9} - F_{2,11})  + (G_{3,9} + G_{1,11})  \nonumber \\
\Xi^*_5 & = & (F_{1,10} + F_{3,12}) + (G_{2,10} - G_{4,12})   \;\;\;\;\;\;\;\;\;\;  \Xi^*_6   =  (F_{2,10} + F_{4,12}) -  (G_{1,10} - G_{3,12})  \nonumber \\
\Xi^*_7 & = & (F_{3,10} - F_{1,12}) - (G_{4,10} + G_{2,12})   \;\;\;\;\;\;\;\;\;\;   \Xi^*_8  =  (F_{4,10} - F_{2,12}) - (G_{3,10} + G_{1,12}) 
\eea
Note that  fermions $\Xi^{*}$'s are obtained from the $\Xi$'s givem by eq.(\ref{eq:Xi}) by complex conjugation, i.e. replacing $G_{j,8+k}$ by $-G_{j,8+k}$.
The $\Xi$'s, as well as the  $\Xi^*$'s, stand in the representation ($1/2, {\bf 4}$) of the algebra ${\it SO(3)_{-}} \oplus {\it Sp(4)}$. It can   also be noted that  ${\it SO(3)_{+}}$ acts trivially on the $\Xi$'s and  $\Xi^*$'s.

As a conclusion, we have determined two (conjugate) super Schr\"odinger algebras in the Lie super-algebra {\it Osp(6,2/4)}, the first one being
\be
{\it \widetilde{Sch}(4/4)_{symp}}  =   [{\it SO(3)}_{-} \oplus  {\it Osp(4^*/4)}]\,  \rhd \,  {\it SH(4/4)}
\ee
and the second one differing from the first simply by the super-Heisenberg part, ${\it SH^*(4/4)}$ replacing {\it SH(4/4)}. As in the previous section, we formally represent the {\it Osp(6,2/4)} superalgebra as:
\be
{\it Osp(6,2/4)} =_ {v.s.}   {\it SH(4/4)} \,  \lhd \,    [{\it Osp(4^*/4)} \oplus {\it SO(3)}_{-} \oplus   {\it SO(1,1)}] \,  \rhd \,  {\it SH^*(4/4)} \label{eq:dosp}
\ee
keeping in mind that the generator $M$ in {\it SH(4/4)} and $M^*$ in ${\it SH^*(4/4)}$ close under commutation relations into the dilatation generator $\Lambda$, represented in {\it O(1,1)} part of the decomposition eq.(\ref{eq:dosp}). We also remind that the bosonic part of ${\it Osp(4^*/4)}$ is 
${\it SO(2,1)}_c \oplus {\it SO(3)}_{+} \oplus   {\it  Sp(4)}$ where {\it  Sp(4)} is compact,  ${\it SO(2,1)}_c$ is generated by $H, C, D$ and  ${\it SO(3)}_{+}$ together with ${\it SO(3)}_{-}$ form the  `rotation"  algebra acting on the $P_{a}$'s and   $K_{a}$'s (respectively $P^*_{a}$'s and   $K^*_{a}$'s) that is
\be  
{\it SO(4)} = {\it SO(3)}_{+} \oplus {\it SO(3)}_{-}
\ee
The generalisation to {\it Osp(6,2/2N)} is straightforward, the compact  {\it  Sp(4)} algebra being replaced by {\it  Sp(2N)}, $N$ positive integer, and one gets
\be
{\it \widetilde{Sch}(2N/4)_{symp}}  =   [{\it SO(3)}_{-} \oplus  {\it Osp(4^*/2N)}]\,  \rhd \,  {\it SH(4/2N)}
\ee
and
\be
{\it Osp(6,2/N)} =_{v.s.}    {\it SH(4/2N)} \,  \lhd \,    [{\it Osp(4^*/2N)} \oplus {\it SO(3)}_{-} \oplus   {\it SO(1,1)}] \,  \rhd \,  {\it SH^*(4/2N)} 
\ee

  \section{The case of {\it F(4;2)}}
  
 As it could be expected, the super Schr\"odinger  algebra which can be extracted from {\it F(4;2)} exhibits some exceptional features. In particular the supersymmetric extension of the Heisenberg part {\it H(3)}, arising from {\it SO(5,2)}, will not be obtained by adding a triplet but a quadruplet of fermions, in other words a spinorial representation, namely labelled by $j = 3/2$, of the associated rotation group {\it SO(3)}. 
 
 In order to determine the real form of the  {\it F(4)} superalgebra we are interested in, that is the one with 
{\it SO(5,2)} $ \oplus $ {\it SU(2)} as bosonic part, let us first start by considering the {\it F(4)}  superalgebra defined on the complex field {\bf C} \cite{FSS}. Then, its bosonic part  {\it Sl(2)} $ \oplus $ {\it SO(7)} can be generated by the elements  $T_i$ ($i=1,2,3$) and $M_{pq} = -M_{qp}$ ($p,q=1,\ldots,7$) respectively satisfying
 \bea
&& [T_i, \, T_j] =  i \vr_{ijk} \, T_k\;\;\;\;\;\;\;\;\;\;\;\;\;\;\; \;\;\;\; [T_i, \, M_{pq}] =  0 \\
&& [ M_{pq}, \, M_{rs}]  = \delta_{qr} \, M_{ps}  + \delta_{ps} \, M_{qr}  - \delta_{pr} \, M_{qs} - \delta_{qs} \, M_{pr} 
\eea
while the $16$ generators of its fermionic part stand in the representation $({\bf 2},{\bf 8}) \equiv (1/2;1/2,1/2,1/2)$ of {\it Sl(2)} $ \oplus $ {\it SO(7)}. We denote by $F_{\alpha \mu}$, $\alpha = \pm;$  and $\mu =(\pm,\pm.\pm)$ a basis of generators satisfying the  relations
\bea
&& [T_i, \, F_{\alpha, \mu}] = \frac{1}{2} \, \sigma^{i}_{\beta \alpha} \,  F_{\beta, \mu} \;\;\;\;\;  \;\;\;\;\; \ \
[M_{pq}, \, F_{\alpha, \mu}] = \frac{1}{2} \, (\Gamma_p \Gamma_q)_{\nu \mu} \,  F_{\alpha, \nu} \\
&& \{F_{\alpha, \mu}, \, F_{\beta, \nu} \} =  2 \, C^{(8)}_{\mu \nu} \, (C^{(2)} \sigma^i)_{\alpha \beta} \, T_i
+  \frac{1}{3} \,  C^{(2)}_{\alpha \beta} \, (C^{(8)} \Gamma^p \Gamma^q)_{\mu \nu} \, M_{pq}
 \eea
The $\sigma^i$ are the usual Pauli matrices.  The $8$-dim. matrices   $\Gamma_{p}$ form a Clifford algebra
 \be
 \{\Gamma_{p}, \, \Gamma_{q}\} = 2 \, \delta_{pq} 
 \ee
and are chosen as
\bea
&& \Gamma_{1} =  \sigma_1 \otimes \sigma_1 \otimes \sigma_1  \;\;\;\;\   \Gamma_{2} =  \sigma_1 \otimes \sigma_1 \otimes \sigma_2  \;\;\;\;\  \Gamma_{3} =  \sigma_1 \otimes \sigma_1 \otimes \sigma_3  \;\;\;\;\; \;\;\Gamma_{4} =  \sigma_1 \otimes \sigma_2 \otimes {\bf 1}  \nonumber \\
&&   \Gamma_{5} =  \sigma_1 \otimes \sigma_3 \otimes {\bf 1} \;  \;\;\;\; \;\; \Gamma_{6} =  \sigma_2 \otimes  {\bf 1}  \otimes   {\bf 1} \;\;\;\;\; \;\;\;
  \Gamma_{7} =  \sigma_3 \otimes  {\bf 1}  \otimes   {\bf 1}  \label{eq:gamma8}
\eea
We note that $ \overline{\Gamma_{p}} = (-1)^{p + 1} \Gamma_{p}$ and that  $\Gamma_{p} \Gamma_{q}$ generate the Lie algebra  {\it SO(7)}.  Finally  $C^{(2)} $ and $C^{(8)}$ are, respectively, the $2 \times 2$ and $8 \times 8$ real charge conjugation matrices given by 
\be
C^{(2)} = i  \sigma_{2} \;\;\;\;\;\;\;\;\;\ \;\;\;\;\;\;\;\;\;\ C^{(8)} =  \Gamma_{1} \Gamma_{3}  \Gamma_{5} \Gamma_{7}  = (i \sigma_2) \otimes \sigma_3 \otimes (i\sigma_2)
\ee
Following \cite{Parker} the {\it F(4)} real form with  {\it SO(5,2)} $ \oplus $ {\it SU(2)} as bosonic part is made from the elements $ X + {\it C}_0(X)$, $X$ belonging to the (complex) {\it F(4)}  superalgebra
and $ {\it C}_0$ being the semi-involutive semi-morphism of {\it F(4)} acting on the bosonic part as follows ($ X^{\dag} = \overline{X}^t$)
\bea
{\it C}_0(X) & = & - X^{\dag} \;\;\;\;\ \;\;\;\;\; X \in {\it Sl(2)}  \nonumber \\
{\it C}_0(X) & = & {\it  \tau}(X) \;\;\;\;\ \;\;\;\;   X \in {\it SO(7)}  
\eea
with  ${\it  \tau}$ acting on the orthogonal algebra generated by   $\Gamma_{p} \Gamma_{q}$ ($ 1 \leq p \leq q \leq 7$) as
\be
{\it  \tau}(\Gamma_{p} \Gamma_{q}) =  \Gamma_4 \, \overline{\Gamma_{p}} \overline{\Gamma_{q}} \,   \Gamma_4^{-1}  
\ee
and acting on the fermionic part as follows 
\be
{\it C}_0({\bf v} \otimes  {\bf x}) = i J {\bf v} \otimes  \Gamma_4 \, {\bar{\bf x}}
\ee
where ${\bf v} \otimes  {\bf x}$ is the most general element of the representation $(1/2) \otimes (1/2,1/2,1/2)$ and $J$ acts on the states  $F(\pm)$ of the {\it Sl(2)} $2$-dim. representation as
\be
J \, F(\pm) = \pm \, F(\pm)
\ee
In this framework a basis of the  {\it SO(5,2)} algebra shows up made by the  $11$ elements  of the maximal compact subalgebra {\it SO(5) $ \oplus $ SO(2)}
\be
 \{M_{13}, M_{14}, M_{15}, M_{17}, M_{34}, M_{35}, M_{37}, M_{45}, M_{47}, M_{57} \} \otimes  \{M_{26} \} \equiv {\it SO(5)}  \otimes {\it SO(2)}  \label{eq:so5,2}  
 \ee
and by the $10$ non-compact generators 
\be
i \{M_{12}, M_{16}, M_{23}, M_{24}, M_{25}, M_{27}, M_{36}, M_{46}, M_{56}, M_{67} \}
\ee
while for  the  {\it SU(2)} remaining algebra - the $R$-symmetry of our problem - one gets the generators $i T_j$ ($i=1,2,3$).

As for the fermionic part of {\it F(4;2)}, it can be seen as the direct sum of two ($1/2, 1/2$) representations of the   {\it SO(3)} $\oplus$  {\it SU(2)} algebra, where the considered  {\it SO(3)} algebra is generated by
\be
{\it SO(3)}  \equiv   \{M_{13},  M_{14},  M_{34} \}  \label{eq:rot}
\ee
and the {\it SU(2)} part by the $i T_j$ generators. This  {\it SO(3)} algebra will be chosen (see below) as 
the rotation algebra acting on $P_j$ and $ K_j$ ($j = 1,2,3$) elements of our  Heisenberg algebra: that is why we prefer to keep the notations   {\it SO(3)}, instead of {\it SU(2)}, for this algebra, although it acts on spinorial representations. A basis for these two representation reads: 
\bea
&&   F(+;  +,+,+) - F(-; -, -,+) \;\;\;\;\; \;\;\;\;\; \;  i(F(+;  +,+,+) + F(-; -, -,+))    \nonumber \\
&& F(+;  +,+,-) - F(-; -, -,-) \;\;\;\;\; \;\;\;\;\; \;  i(F(+;  +,+,-) + F(-; -, -,-)) \nonumber  \\
&& F(-; +, +,+) +  F(+;  -,-,+) \;\;\;\;\; \;\;\;\;\; \;   i(F(-;  +,+,+) - F(+; -, -,+)) \nonumber    \\
&&  F(-; +, +,-) + F(+;  -,-,-)  \;\;\;\;\; \;\;\;\;\; \;  i(F(-;  +,+,-) - F(+; -, -,-)) 
\eea
for the first one and
\bea
 &&   F(+; +,-,+) + F(-; -, +.+) \;\;\;\;\; \;\;\;\;\; \;  i(F(+;  +,-,+) - F(-; -, +,+))    \nonumber \\
&& F(+;  +,-,-) + F(-; -, +,-)  \;\;\;\;\; \;\;\;\;\; \;  i(F(+;  +,-,-) - F(-; -, +,-)) \nonumber  \\
&&  F(-;  +,-,+) - F(+; -, +,+) \;\;\;\;\; \;\;\;\;\; \;  i(F(-;  +,-,+) + F(+; -, +,+)) \nonumber    \\
&&  F(-;  +,-,-) - F(+; -, +,-) \;\;\;\;\; \;\;\;\;\; \;  i(F(-;  +,-,-) + F(+; -, +,-))
\eea  
for the second one.

Coming back to the bosonic part, it is convenient to take as Lorentz algebra in the conformal {\it SO(5,2)}  one the {\it SO(4,1)}  algebra generated by the elements $M_{\mu \nu}$ and $M_{\mu 6}$, with $\mu, \nu = 1,3,4,5$, and to consider the embedding
\be
{\it SO(5,2)} \supset  {\it SO(3)} \oplus  {\it SO(2,2)} \cong {\it SO(3)} \oplus  {\it SO(2,1)} \oplus  {\it SO(2,1)'}
\ee
the   ``rotation" {\it SO(3)} algebra being given by eq.(\ref{eq:rot}) and the two {\it SO(2,1)}  by
 \bea
 {\it SO(2,1)} &  \equiv &  \{i(M_{25} + M_{76}), i(M_{56} + M_{72}), M_{57} + M_{26}  \} \\
{\it SO(2,1)'} &  \equiv &  \{ i(M_{25} - M_{76}), i(M_{65} + M_{72}), M_{57} - M_{26} \}
\eea 
Taking {\it SO(2,1)'}  as the  ``non relativistic conformal algebra", we define
 \bea
  H & =  &(M_{57} - M_{26}) -  i(M_{25} -  M_{76})   \label{eq:calF-1} \\ \label{eq:calF-1}
  C & =& -(M_{57} - M_{26}) -  i(M_{25} -  M_{76})  \label{eq:calF-2}  \\ \label{eq:calF-2}
 D & = &  i (M_{72} - M_{56})  \label{eq:calF}
 \eea
 which satisfy the commutation relations
 \be
 [D,\, C]  =  2 C  \;\;\;\;\; \;\;\;\;\;  \; \; [D, \, H] = - 2 H \;\;\;\;\;\; \;\;\;\;\; \;  \; [H, \, C] = -4  D \label{eq:cg}
 \ee
 while the  {\it SO(2,1)}  will contain the elements
 \bea
  M & =  &(M_{57} + M_{26}) -  i(M_{25} +  M_{76})    \\
  M^* & =& (M_{57} + M_{26}) +  i(M_{25} + M_{76})   \\
 	\Lambda & = &   i(M_{27} + M_{65})  \label{eq:so21-F}
 \eea
 satisfying
 \be
 [\Lambda, \, M]  =  2 M  \;\;\;\;\; \;\;\;\;\;   [\Lambda, \, M^*] = - 2 M^* \;\;\;\;\; \;\;\;\;\;  [M, \, M^*] = 4  \Lambda
\ee
Then the $3$-dim Heisenberg algebra {\it H(3)}  contains, in addition to $M$, the elements 
\bea
&& P_1 = iM_{21}  + M_{71} \;\;\;\;\; \;\;\;\;\; \; P_2 = iM_{23}  + M_{73}  \;\;\;\;\; \;\;\;\;\;  P_3
 = iM_{24}  + M_{74}  \\
  && K_1 = iM_{61}  + M_{51} \;\;\;\;\; \;\;\;\;\; K_2 = iM_{63}  + M_{53}  \;\;\;\;\; \;\;\;\; \, K_3
 = iM_{64} + M_{54} 
 \eea
  and the $3$-dim Heisenberg algebra ${\it H^*(3)} $ will be constituted by $M^*$ and
 \bea
 && P^*_1 = iM_{61}  - M_{51} \;\;\;\;\; \;\;\;\;\; P^*_2 = iM_{63}  - M_{53}  \;\;\;\;\; \;\;\;\; P^*_3
 = iM_{64}  - M_{54}  \\
  && K^*_1 = M_{71} - iM_{21}  \;\;\;\;\; \;\;\;\; K^*_2 =M_{73} -  iM_{23}     \;\;\;\;\; \;\;\;\; K^*_3
 = M_{74}  - iM_{24} 
 \eea
 such that 
  \be
 [P_a, \, K_b] = - \delta_{ab} \, M  \;\;\;\;\; \;\;\;\;\;  \;\; \;[P^*_a, \, K^*_b] = - \delta_{ab} \, M^*  \;\;\;\;\; \;\;\;\;\; (a,b = 1,2,3)
 \ee
 Now, from the expressions given by eq.(\ref{eq:so21-F}) one can deduce which fermionic generators commute with $M$ or $M^*$ or both.
 \begin{itemize}
	\item  There are $8$ fermions  commuting with $M$  and  $M^*$:
  \bea
&& Q_1 \equiv (F(+;  +,+,+) - F(-; -, -,+)) + ( F(-;  +,+,-) +  F(+; -, -,-))   \nonumber \\
&& Q_2 \equiv (F(+;  +,+,-) - F(-; -, -,-)) - (F(+;  -,-,+) + F(-; +, +,+)) \nonumber  \\
&& Q_3 \equiv (F(-;  +,-,+) - F(+; -, +,+)) - (F(+;  +,-,-) + F(-; -, +,-)) \nonumber \\
 && Q_4 \equiv (F(+;  +,-,+) + F(-; -, +,+)) + (F(-;  +,-,-) -  F(+; -, +,-)) \nonumber \\
 && S_1 \equiv i[(F(+;  +,+,+) + F(-; -, -,+)) -  (F(-;  +,+,-) -  F(+; -, -,-))]  \nonumber \\
&& S_2 \equiv i[ (F(+;  +,+,-) + F(-; -, -,-)) - (F(+;  -,-,+) - F(-; +, +,+))] \nonumber  \\
&& S_3 \equiv i[(F(-;  +,-,+) + F(+; -, +,+)) + (F(+;  +,-,-) - F(-; -, +,-) )] \nonumber \\
 && S_4 \equiv  i[(F(+;  +,-,+) - F(-; -, +,+)) - (F(-;  +,-,-) +  F(+; -, +,-))] 
 \eea
\item there are $4$ fermions commuting with $M$  but not with $M^*$:
\bea
&& \Xi_1 =  (F(+;  +,+,+) - F(-; -, -,+)) - (F(-;  +,+,-) +  F(+; -, -,-))   \nonumber \\
&& \Xi_2 = (F(+;  +,+,-) - F(-; -, -,-)) + (F(+;  -,-,+) + F(-; +, +,+)) \nonumber  \\
&& \Xi_3 = i[ (F(+;  +,+,-) + F(-; -, -,-)) + (F(+;  -,-,+)-  F(-; +, +,+))] \nonumber \\
 &&\Xi_4 = i[(F(+;  +,+,+) + F(-; -, -,+)) +  (F(-;  +,+,-) -  F(+; -, -,-))] 
 \eea
\item  and finally $4$ fermions   commuting with $M^*$  but not with $M$:
 \bea
&& \Xi^*_1 =(F(-;  +,-,+) - F(+; -, +,+)) + (F(+;  +,-,-) + F(-; -, +,-)) \nonumber \\
 && \Xi^*_2 = (F(+;  +,-,+) + F(-; -, +,+)) -  (F(-;  +,-,-) -  F(+; -, +,-)) \nonumber \\
&& \Xi^*_3 = i[(F(-;  +,-,+) + F(+; -, +,+)) - (F(+;  +,-,-) - F(-; -, +,-) )] \nonumber \\
 &&\Xi^*_4 =  i[(F(+;  +,-,+) - F(-; -, +,+)) + (F(-;  +,-,-) +  F(+; -, +,-))] 
 \eea
\end{itemize}
We note that the $\Xi_{\mu}$'s as well as  $ \Xi^*_{\mu}$'s ($\mu = 1,2,3,4$)  transform as the representation ($3/2$) under the rotation algebra {\it SO(3)} given by eq.(\ref{eq:rot}), as the transformation $(1/2) \oplus (1/2)$ of the $R$-symmetry algebra {\it SU(2)}, while   the  ``conformal algebra" ${\it SO(2,1)'}$, generated by the elements given by eqs (\ref{eq:calF-1})-(\ref{eq:calF-2})-(\ref{eq:calF}),  acts trivially on each of these elements. The commutation and anti-commutation relations
\be
[P_a, \, \Xi_{\mu}] = [K_a, \, \Xi_{\mu}]  = 0 \;\;\;\;\; \;\;\;\;\; \{\Xi_{\mu}, \, \Xi_{\nu}\} = 4 \, \delta_{\mu \nu} \, M
;\;\;\;\; \;\;\;\;\; a= 1,2,3; \, \mu = 1,2,3,4
\ee
and
\be
[P^*_a, \, \Xi^*_{\mu}] = [K^*_a, \, \Xi^*_{\mu}]  = 0 \;\;\;\;\; \;\;\;\;\;  \{\Xi^*_{\mu}, \, \Xi^*_{\nu}\} = 4 \, \delta_{\mu \nu} \, M^*  \;\;\;\;\; \;\;\;\;\; a= 1,2,3; \, \mu = 1,2,3,4
\ee
allow to consider the super-algebra generated by $\{P_a, K_a,  \Xi_{\mu}\}$ and by $\{P^*_a, K^*_a,  \Xi^*_{\mu}\}$, respectively, as the supersymmetric extensions of the ${\it H(3)} $ and ${\it H^*(3)} $
Heisenberg algebras. Due to the presence of the spinorial representation ($3/2$) in the fermionic sector we will denote them, respectively, as   ``spinorial super-Heisenberg algebra ${\it SH(3/2)}_{spin}$ and 
${\it SH^*(3/2)}_{spin}$". 

Finally, considering the anti-commutation relations of the $8$ fermions $Q_{\mu}$ and $S_{\mu}$, which commute with $M$ and $M^*$, we can reconstruct the  ${\it SO(3)} \oplus  {\it SU(2)} \oplus  {\it SO(2,1)}'$ direct sum.  Moreover,  while the rotation {\it SO(3)} as well as the $R$-symmetry {\it SU(2)} algebras split the fermionic sector into two $4$-dim. representations, namely $\{Q_1, Q_2, S_1, S_2\}$ and $\{Q_3, Q_4, S_3, S_4\}$, the conformal algebra  {\it SO(2,1)}' decomposes it into $4$ $2$-dim. representations, generated by $\{Q_1, Q_3\}$, $\{Q_2, Q_4\}$, $\{S_1, S_3\}$ and $\{S_2, S_4\}$. 

Therefore we recognize the {\it Osp(4/2,R)} algebra built from this set of seventeen elements. One can directly check that this {\it Osp(4/2,R)}  transforms the ${\it SH(3/2)}_{spin}$ and 
${\it SH^*(3/2)}_{spin}$ super-Heisenberg into themselves leading to the conclusion that the  super Schr\"odinger  algebra in {\it F(4;2)} is the semi-direct sum
\be
{\it \widetilde{Sch}(3/2)}_{F_4} \equiv  {\it Osp(4/2,R)}  \, \rhd \,  {\it SH(3/2)}_{spin}
 \ee
 and that  {\it F(4;2)}  can be formally decomposed as:
 \be
{\it F(4;2)}  =_{v.s.} {\it SH(3/2)}_{spin} \, \lhd \, [{\it Osp(4/2,R)}  \, \oplus \, {\it O(1,1)}] \, \rhd \,  {\it SH^*(3/2)}_{spin}
\ee
the  {\it SO(1,1)} algebra commuting with {\it Osp(4/2,R)}  being generated by the generator $\Lambda$,  see eq.(\ref{eq:so21-F}).

\section{Conclusion}

The super-symmetric extension of the  Schr\"odinger  algebra has been constructed in each admissible
super-conformal algebra.  Although the obtained symmetries present some differences according to the type of considered superconformal algebra, they always show up as the semi-direct sum of a unitary or orthosymplectic superalgebra (to which is sometimes added a  {\it SO(2)} or  {\it SO(3)} factor) acting on a super-Heisenberg part, see Table 1:
\begin{table}[htbp]
\label{tablegc}
\begin{center}
\begin{tabular}{|c||c|}
\hline
 Admissible superconformal algebras  & super Schr\"odinger algebras  \\
 \hline 
  \\
 \vspace{1,5mm}
  {\it SU(2,2/N)}  \;\;\;\; $N \neq 4$ & [{\it SO(2)} $ \oplus $  {\it SU(1,1/N)}] $ \rhd $ {\it SH(2/N)} \\ \vspace{1,5mm}
   {\it pSU(2,2/4)}    & {\it SU(1,1/N)} $ \rhd$ {\it SH(2/N)} \\ \vspace{1,5mm}
   {\it Osp(N/4,R)} &  {\it  Osp(N/2,R)} $ \rhd$ {\it SH(1/N)}     \\ \vspace{1,5mm}
   {\it Osp(6,2/2N)} &     $[{\it SO(3)}  \oplus   {\it Osp(4^*/2n)}]  \rhd {\it SH(4/2n)}$   \\ \vspace{1,5mm}
    {\it F(4;2)} &     ${\it Osp(4/2,R)}  \rhd {\it SH(3/2)}_{spin}$ \\
 \hline
\end{tabular}
\caption{super Schr\"odinger symmetries in superconformal algebras.}
\end{center}
\end{table}

 One may note that the super-Heisenberg part constructed from the exceptional {\it F(4;2)} presents a difference with respect to the other cases, its fermionic sector transforming as a spinorial representation of the corresponding {\it SO(3)} rotation algebra. This is, of course, a consequence of the spinorial character of the fermionic  {\it F(4;2)} sector and it is already known that this exceptional superalgebra gives peculiar results \cite{DFV}. We also remark that the super Schr\"odinger  algebras arising from {\it Osp(6,2/2N)} differ from those extracted in the other unitary and orthosymplectic ones for an extra {\it SO(3)}, part of {\it SO(4)} rotation symmetry of the problem, and added to the $ {\it Osp(4^*/2N)}$ factor. Let us at this point mention that this case is the only one with a $R$-symmetry of the symplectic type. Finally, let us emphasize  the property of any 
super Schr\"odinger  algebra to occupy a particular position in its corresponding superconformal algebra, where an adequate basis decomposition provides two copies of the super Schr\"odinger  
symmetry, differing by their super-Heisenberg parts.

We hope that this study and the characterization of the super Schr\"odinger  algebras that we have obtained will help to develop properties of such symmetries, for ex. the theory of their representations \footnote{Representations of the  Schr\"odinger group have been constructed in \cite{Perr}.}, and mainly to better understand the physics behind their structures. 

\bigskip

\bigskip

\noindent {\bf  \large Acknowledgements}: P.S. is indebted to L. Alvarez-Gaum\'e and S. Ferrara for discussions and useful informations. He wishes also to thank the CERN Theory Division where this work started.

\appendix
\section*{Appendix: Anti-commutation relations of the fermionic generators of ${\it Osp(6,2/4)}$.}
 
   The anti-commutation relations of the fermionic generators  \{$F;G$\} defined by eqs.(\ref{eq:F})-(\ref{eq:G}) are:
\bea
\{F_{i,9}, \, F_{j,9} \} & = & \{F_{i,11}, \, F_{j,11} \} = (E_{i,4+j} - E_{4+j,i}) + (E_{j,4+i} - E_{4+i,j}) \nonumber \\
& & + 2 \, \delta_{ij} (E_{11,9} - E_{9,11}) \\
\{F_{i,9}, \, F_{j,10} \} & = & \{F_{i,11}, \, F_{j,12} \} =  \delta_{ij} (E_{11,10} - E_{10,11} - E_{9,12} + E_{12,9})   \\
\{F_{i,9}, \, F_{j,11} \} & = & \{F_{i,10}, \, F_{j,12} \} = -(E_{i,j} - E_{j,i}) + (E_{4+j,4+i} - E_{4+i,j4+j}) \\
\{F_{i,9}, \, F_{j,12} \} & = & - \{F_{i,10}, \, F_{j,11} \} =  \delta_{ij} (E_{9,10} - E_{10,9} + E_{11,12} - E_{12,11}) \\
\{F_{i,10}, \, F_{j,10} \} & = & \{F_{i,12}, \, F_{j,12} \} = (E_{i,4+j} - E_{4+j,i}) + (E_{j,4+i} - E_{4+i,j}) \nonumber \\
& & + 2  \, \delta_{ij} (E_{12,10} - E_{10,12})  \\
\{G_{i,9}, \, G_{j,9} \} & = & \{G_{i,11}, \, G_{j,11} \} = E_{i,4+j} + E_{4+j,i}  - E_{4+i,j} -  E_{j,4+i}  \nonumber \\
& & + 2 \, \delta_{ij} (E_{9,11} - E_{11,9}) \\
\{G_{i,9}, \, G_{j,10} \} & = & \{G_{i,11}, \, G_{j,12} \} =  \delta_{ij} (E_{10,11} - E_{11,10} + E_{9,12} - E_{12,9})   \\
\{G_{i,9}, \, G_{j,11} \} & = & \{G_{i,10}, \, G_{j,12} \} = (E_{i,j} -  E_{j,i} )+ (E_{4+j,4+i} - E_{4+i,j4+j})\\
\{G_{i,9}, \, G_{j,12} \} & = & - \{G_{i,10}, \, G_{j,11} \} =  \delta_{ij} (-E_{9,10} + E_{10,9} - E_{11,12} + E_{12,11}) \\
\{G_{i,10}, \, G_{j,10} \} & = & \{G_{i,12}, \, G_{j,12} \} = E_{i,4+j} + E_{4+j,i}) - E_{j,4+i} - E_{4+i,j} \nonumber \\
& & + 2  \, \delta_{ij} (E_{10,12} - E_{12,10})  \\
\{F_{i,9}, \, G_{j,9} \} & = &  i \, [(E_{j,4+i} - E_{4+i,j}) - (E_{i,4+j} - E_{4+j,i})  \nonumber \\
& & + 2 \, \delta_{ij} (E_{9,11} + E_{11,9})] \\
\{F_{i,9}, \, G_{j,10} \} & = & \{F_{i,10}, \, G_{j,9} \} =  -\{F_{i,11}, \, G_{j,12} \}   = i \, \delta_{ij} (E_{9,12} + E_{12,9} + E_{10,11} + E_{11,10}) \\
\{F_{i,9}, \, G_{j,11} \} & = &i \, [  -(E_{i,j} - E_{j,i}) + (E_{4+i,4+j} - E_{4+j,4+i})  - 2 \, \delta_{ij} (E_{9,9} - E_{11,11})\ \\
\{F_{i,9}, \, G_{j,12} \} & = & \{F_{i,10}, \, G_{j,11} \} = \{F_{i,11}, \, G_{j,10} \} = \{F_{i,10}, \, G_{j,11} \} = 
\{F_{i,12}, \, G_{j,9} \} =\nonumber \\ &&  i \, \delta_{ij} (-E_{9,10} - E_{10,9} + E_{11,12} + E_{12,11}) \\
\{F_{i,10}, \, G_{j,10} \} & = &  i \, [ (E_{j,4+i} - E_{4+i,j}) - (E_{i,4+j} - E_{4+j,i}) +   2  \, \delta_{ij} (E_{10,12} + E_{12,10})] \\
\{F_{i,10}, \, G_{j,12} \} & = &  i \, [ -(E_{i,j} - E_{j,i}) + (E_{4+i,4+j} - E_{4+j,4+i}) +   2  \, \delta_{ij} (E_{12,12} - E_{10,10})] \\
\{F_{i,11}, \, G_{j,9} \} & = &  i \, [(E_{i,j} - E_{j,i}) - (E_{4+i,4+j} - E_{4+j,4+i}) -  2  \, \delta_{ij} (E_{9,9} - E_{11,11})]  \\
\{F_{i,11}, \, G_{j,11} \} & = &  i \, [-(E_{i,4+j} - E_{4+j,i}) - (E_{4+i,j} - E_{j,4+i}) -  2  \, \delta_{ij} (E_{9,11} + E_{11,9})] \\
\{F_{i,11}, \, G_{j,10} \} & = & = \{F_{i,12}, \, G_{j,9} \} =  i \, (-E_{9,10} - E_{10,9} + E_{11,12} + E_{12,11}) \\
\{F_{i,12}, \, G_{j,10} \} & = &  i \, [ (E_{i,j} - E_{j,i}) - (E_{4+i,4+j} - E_{4+j,4+i}) - 2  \, \delta_{ij} (E_{10,10} -E_{12,12})] \\
\{F_{i,12}, \, G_{j,12} \} & = &  i \, [ -(E_{i,4+j} - E_{4+j,i}) - (E_{4+i,j} - E_{j,4+i}) - 2  \, \delta_{ij} (E_{10,12} + E_{12,10})] 
\eea

 \end{document}